\documentclass[fleqn,usenatbib]{mnras}

\usepackage{newtxtext,newtxmath}

\usepackage[T1]{fontenc}
\DeclareRobustCommand{\VAN}[3]{#2}
\let\VANthebibliography\thebibliography
\def\thebibliography{\DeclareRobustCommand{\VAN}[3]{##3}\VANthebibliography}

\usepackage{graphicx}	
\usepackage{amsmath}	
\usepackage[dvipsnames]{color}
\usepackage[normalem]{ulem}
\usepackage{threeparttable}
\usepackage{multirow}
\usepackage{array}

\newcommand{\OIII}{[O~{\sc iii}]}

\newcommand{\HII}{H~{\sc ii}\ }

\newcommand{\HI}{H~{\sc i}\ }
\newcommand{\Ha}{H$\alpha$\ }
\newcommand{\Hb}{H$\beta$\ }
\newcommand{\kms}{\,\mbox{km}\,\mbox{s}^{-1}}

\newcommand{\SIIHa}{[S~{\sc ii}]/H$\alpha$\ } 
\newcommand{\NIIHa}{[N~{\sc ii}]/H$\alpha$\ }
\newcommand{\OIIIHb}{[O~{\sc iii}]/H$\beta$\ }
\newcommand{\be}{\begin{equation}}
\newcommand{\ee}{\end{equation}}

\definecolor{violet}{rgb}{0.8,0,1}

\def\revone{\textbf}
\def\revone{}

\newcommand{\mainproperties}{
\begin{table}
    \centering
    \caption{General parameters of the galaxy Sextans B}
    \begin{tabular}{lr}\hline
         Parameter&Value  \\ \hline
         Distance$^b$& $1.39 \pm 0.04$ \revone{Mpc}\\ 
         $M_{\rm{B}}^a$ & $-14.07^m$\\
         Linear scale & 6.7~pc~arcsec$^{-1}$\\
         Optical radius$^c$, $R_{25}$ & 147~arcsec = 985~pc \\ 
        $\log(SFR_{FUV})^d$ &  \revone{$-2.44\ \mathrm{M_\odot\ yr^{-1}}$}\\
        12 + log(O/H)$^e$ & \revone{$7.84 \pm 0.05$}\\ 
        $M_{\rm{HI}}^f$ & $4.07 \times 10^7 M_\odot$\\
        $M_{\rm{*}}^g$ & $16.52\pm 13.3\times 10^7 M_\odot$\\
        Scale height$^h$ & $639\pm224$~pc\\
        Kinematic parameters$^i$:&\\
        RA (J2000.0) & $09:59:59.9$\\
        DEC (J2000.0) & $05:19:57$\\
        $V^{\rm{sys}}_{\rm{radio}}\revone{(heliocentric)}$ & $302 \pm 0.9$ km~s$^{-1}$\\
        Inclination, $i$ & $49^\circ$\\
        $PA_{\rm{kin}}$ & $56.6^\circ$\\
        $V_{\rm{rot}}$ at 50 arcsec & $13.2$ km~s$^{-1}$\\
        \hline
    \end{tabular}
    
\begin{tablenotes}
    \scriptsize
    \item $^a$ LV galaxy database \\ \citep[][\url{https://www.sao.ru/lv/lvgdb/}]{Karachentsev2004}
    \item $^b$ \cite{Dalcanton2009}; $^c$ HyperLEDA Database\\ \citep[][\url{http://http://leda.univ-lyon1.fr/}]{Makarov2014}
    \item $^d$ \cite{Hunter2010}; $^e$ \cite{Kniazev2005}; $^f$\cite{Hunter2012}; 
    \item $^g$ \cite{Weisz2011}; $^h$ \cite{Stilp2013}; $^i$ \cite{Namumba2018}
\end{tablenotes}
    \label{tab:main_properties}
\end{table}}

\newcommand{\observations}{
\begin{table*}
	\caption{Log of observational data}
	\label{tab:observations}
	\centering
	{
	\begin{tabular}{llrlcclll}
		\hline
		\multirow{2}{*}{Data set}       & \multirow{2}{*}{Date of obs.}    & \multirow{2}{*}{$\mathrm{T_{exp}}$, s} & \multirow{2}{*}{FOV}                             & pixel size,             & seeing,   & \multirow{2}{*}{sp. range}          & \multirow{2}{*}{$\delta\lambda$, \AA}        \\ 
        &&&&arcsec&arcsec&&\\
        \hline
		FPI/Scorpio-2/BTA  & 2018 Feb {07} & $40\times180$  & {$6.1\arcmin\times6.1\arcmin$} & {0.71} & 2.0              & {8.8~\AA\, around \Ha}  &{0.48} \\
		RSS/SALT PA=120.2 & 2019 May 27 & 2150 & {$1.5\arcsec\times8\arcmin$} & {0.25} & 1.7 & {4347--7369}  & $4.80$ \\
		RSS/SALT PA=159.8 & 2019 May 24 & $2\times1140$ & {$1.5\arcsec\times8\arcmin$} & {0.25} & 1.2 & {4347--7369}  & $4.80$ \\
  		TDS/CMO PA=-6 & 2023 Apr 23 & $1200\times4$ & {$1.5\arcsec\times3\arcmin$} & {0.35} & 1.7 & {3600–5770 \& 5670–7460}  & $2.6\ \&\ 2.4$ \\
		\hline
  
	\end{tabular}}
	\begin{tablenotes}
	\footnotesize
	\item {$\mathrm{T_{exp}}$ is the exposure time;}
	 {FOV is the field of view;}
	 
	\item {seeing is the final angular resolution;} 
	 {$\delta\lambda$ is the final spectral resolution.}
	\end{tablenotes}
\end{table*}
}

\newcommand{\bubbleparams}{
\begin{table*}
    \caption{Properties of the identified expanding ionised superbubbles in Sextans~B.}
    \centering
    {
    \begin{tabular}{cccccccccccccc}
    \hline
         \multirow{2}{*}{\#}&\multirow{2}{*}{RA (J2000)}&\multirow{2}{*}{DEC (J2000)}&$\mathrm{R_a}$,&$\mathrm{R_b}$,&PA,&$\mathrm{R_{eff}}$,&$\mathrm{t_{kin}}$,&$\mathrm{V_{exp}}$,&$\mathrm{L_{mech}}$,&$\mathrm{E_{kin}}$,&$\mathrm{n_0},$&N, \\
         &&&$^{\prime\prime}$&$^{\prime\prime}$&$^\circ$&pc&Myr&$\kms$&$ 10^{36}\ \mathrm{erg\ s}^{-1}$ &$ 10^{51}$ erg&$\mathrm{cm^{-3}}$&O5V \\  \hline
$S 1$&$9:59:58.7$&$5:20:13.5$&    10&     9&$ 90$&$    30$&$  0.41$&$     45$&$     8.7$&$     0.11$&$  0.27$&$   32$ \\
$S 2$&$10:00:01.5$&$5:19:20.7$&     9&     8&$ 87$&$    28$&$  0.33$&$    50$&$     9.5$&$     0.10$&$  0.24$&$   34$ \\
$S 3$&$10:00:05.0$&$5:19:15.9$&    12&    17&$ 50$&$    46$&$  1.05$&$    26$&$     3.5$&$     0.12$&$  0.24$&$   12$ \\

    \hline
    \end{tabular}}
    \label{tab:bubble_params}
\end{table*}
}

\newcommand{\distproperties}
{
\begin{table*}
\caption{Parameters of the \Ha profiles shown in Fig.~\ref{fig:profs}}
    {

\begin{tabular}{cccccccc}\hline
\#& Name & RA (J2000) & DEC (J2000) &\multicolumn{2}{c}{First comp.}&\multicolumn{2}{c}{Second comp.} \\
 &  &&& $\mathrm{V}_1,\ \kms$ & $\mathrm{\sigma}_1,\ \kms$ & $\mathrm{V}_2,\ \kms$ & $\mathrm{\sigma}_2,\ \kms$ \\ \hline
1 & SHK 10 & 10:00:02.9 & 5:20:24.0  & 25 & 14 &  &  \\
2 & SHK 5 & 10:00:00.9 & 5:20:26.3  & 13 & 17 &  &  \\
3 & SHK 4 & 10:00:00.1 & 5:20:14.5  & 7 & 19 &  &  \\
4 & S 1 & 9:59:58.6 & 5:20:13.9  & -78 & 21 & 11 & 31 \\
5 & SHK 2 & 9:59:58.3 & 5:19:45.8  & 18 & 19 &  &  \\
6 & SHK 1 & 9:59:57.9 & 5:19:40.3  & 12 & 19 &  &  \\
7 & P 2 & 9:59:58.3 & 5:19:18.5  & -7 & 24 & 46 & 25 \\
8 & SHK 6 & 10:00:00.9 & 5:19:36.5  & 12 & 15 &  &  \\
9 & SHK 8 & 10:00:01.5 & 5:18:45.2  & 13 & 20 &  &  \\
10 & P 3 & 10:00:05.6 & 5:18:02.0  & 23 & 22 &  &  \\
11 & S 2 & 10:00:01.4 & 5:19:19.6  & -77 & 32 & 24 & 23 \\
12 & S 3 & 10:00:05.1 & 5:19:15.8  & -47 & 15 & 5 & 25 \\
13 & P 1 & 10:00:02.3 & 5:19:39.9  & 18 & 51 & 28 & 22 \\ \hline
\end{tabular}}
\begin{tablenotes}
\footnotesize
\item \revone{
HII region names correspond to \cite{Strobel1991} notation.}
\end{tablenotes}

    \label{tab:distproperties}
\end{table*}

}
\newcommand{\RGBHaFUVHI}{
\begin{figure}
    \centering
    \includegraphics[width=\linewidth]{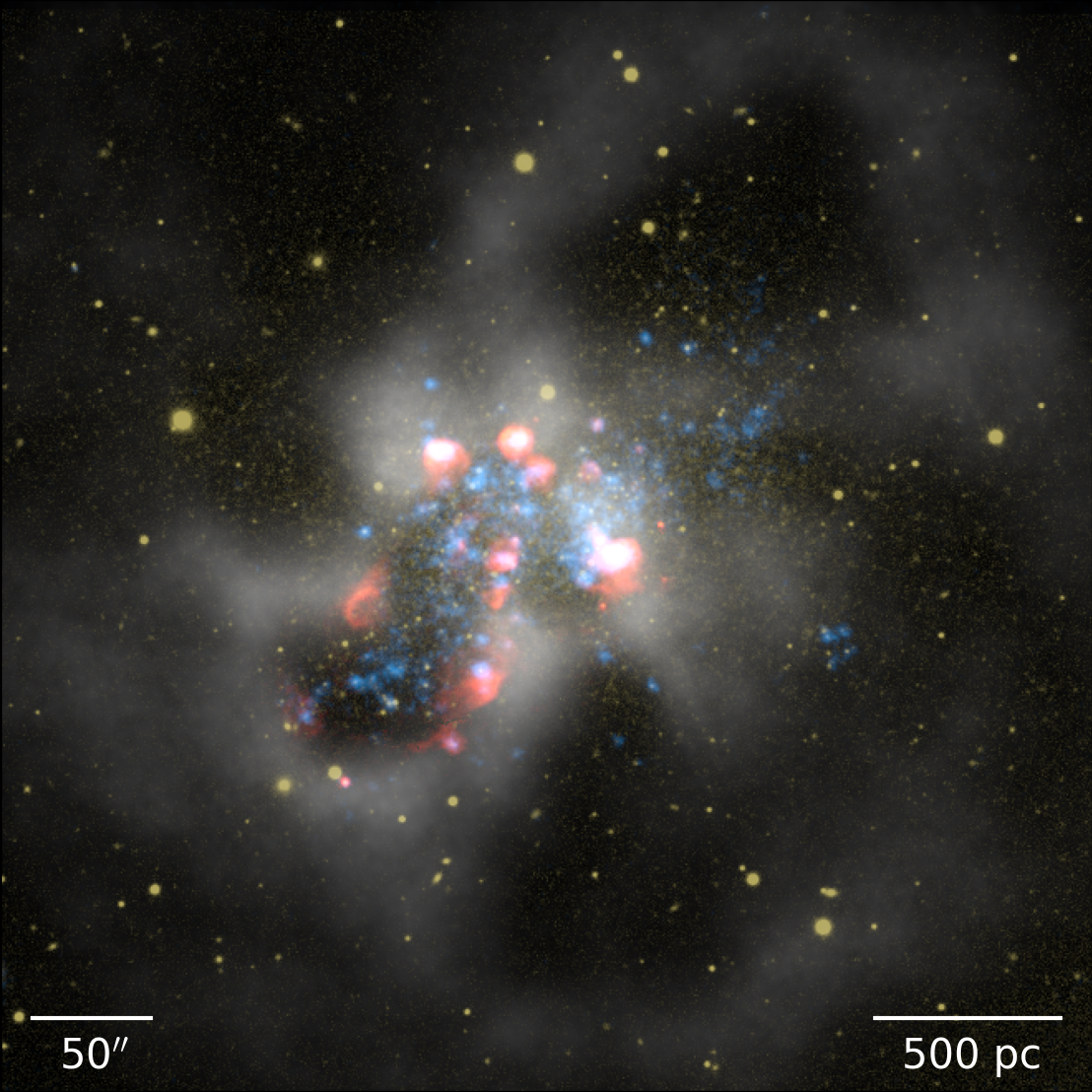}
    \caption[]{\revone{False-colour image\footnotemark{} of the galaxy Sextans~B. The red and blue colours show the distribution of the current and recent star formation activity traced by the \Ha (BTA/FPI; this work) and FUV (GALEX; \citealt{Hunter2010}) emission, respectively. The distribution of the atomic hydrogen \HI 21~cm emission (VLA, LITTLE THINGS survey;  \citealt{Hunter2012}) is shown by white colour. The yellow colour corresponds to the medium-band optical image centred on the continuum near \Ha line (the KPNO telescope; \citealt{Masseylines}).}}
    \label{fig:map}
\end{figure}
}

\newcommand{\velocitymap}{
\begin{figure*}
    \centering
    \includegraphics[width=0.95\linewidth]{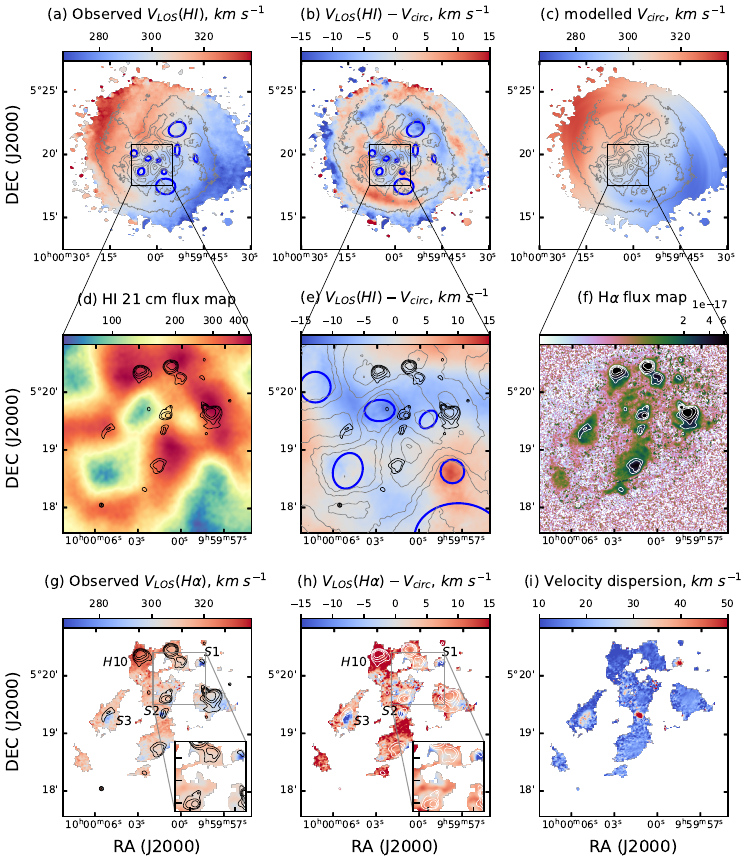}
\caption{Line-of-sight velocity fields in \HI and H$\alpha$, and results of their analysis. Panels (a) and (g) show the observed \HI and \Ha velocity field, respectively. Panel (c) shows the constructed circular rotation model (see Sec.~\ref{sec:velocities}), which was subtracted from the data. The residuals are given in Panels (b), (e) (for H~\textsc{i}) and (h) (for H$\alpha$). Panels (d) and (f) demonstrate the \HI 21~cm \Ha brightness distribution, respectively, shown for the central area inside the black square marked at the panels from the first row. Panel (i) shows the measured intrinsic \Ha velocity dispersion. Black and white contours are lines of constant \Ha brightness, while grey contours show \HI 21~cm constant surface brightness levels. The inset images in Panels (g) and (h) show the \Ha velocity field obtained from the MUSE data (observed and corrected for the circular rotation, respectively). Blue ellipses in Panels (a), (b), (c) and (e) show the localisation of the \HI shells as identified by \citet{Pokhrel2020}.}
    \label{fig:velocitymap}
\end{figure*}

}

\newcommand{\spectrumpaotos}{
\begin{figure*}
    \centering
    \includegraphics[width=.45\linewidth]{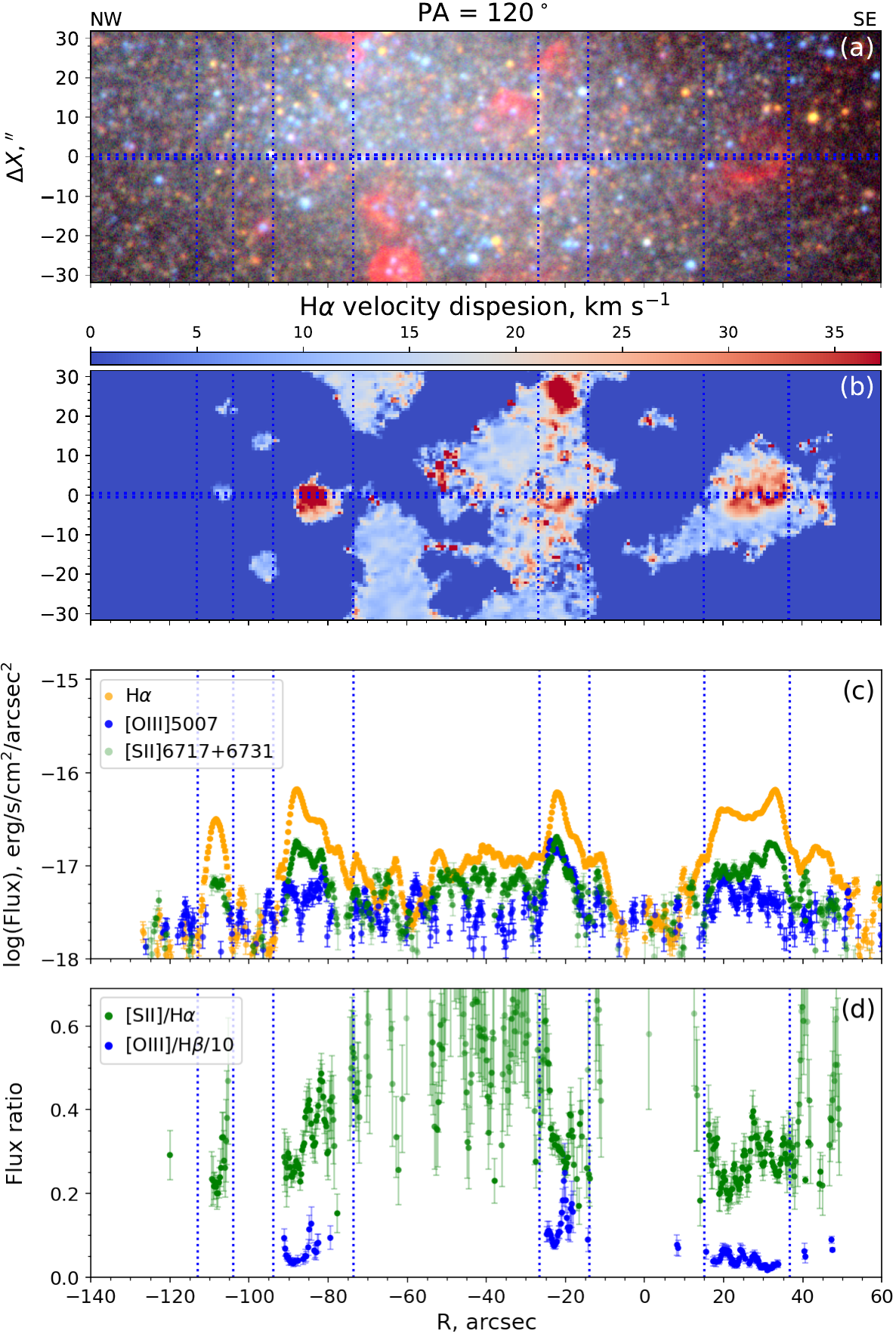}
    ~\includegraphics[width=.45\linewidth]{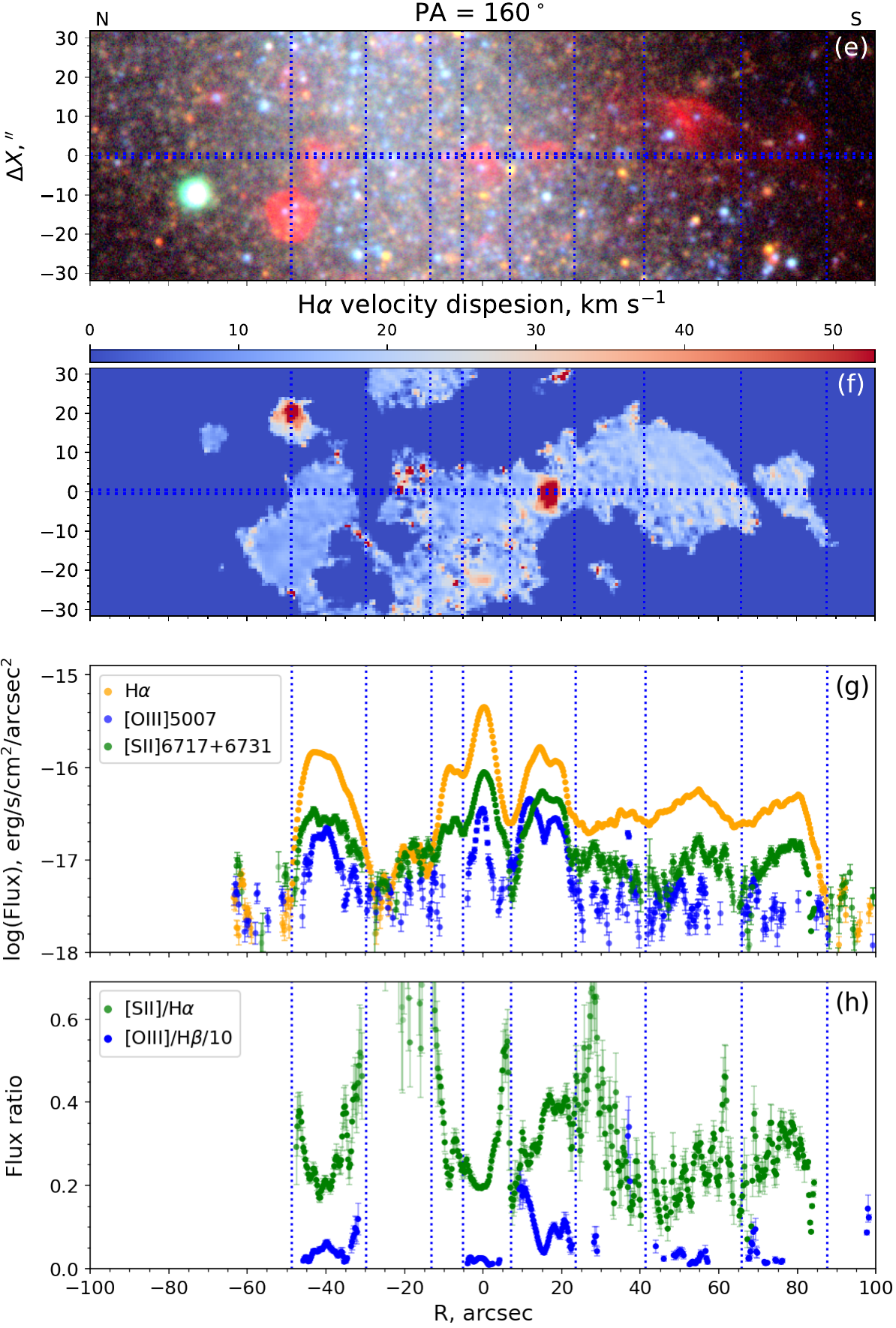}
    
    \caption{\revone{Plots correspond to $\rm PA = 120^\circ$ and $\rm PA = 160^\circ$ for the SALT spectra (left-hand and right-hand panels, respectively). Panels (c) and (g) show the distribution of the emission line fluxes, while on panels (d) and (h) we plot their ratios along the slits. \revone{Panels (a) and (d)} show the RGB image of the galaxy with red channel corresponding to the H$\alpha$ brightness, and blue and green channels -- to the stellar continuum (in V and B bands, respectively, from the KPNO telescope; \citealt{Masseyubvri}). Panels (b) and (e) demonstrate the distribution of the measured intrinsic \Ha velocity dispersion. Localisation of the slit is shown with a horizontal blue dotted line. Vertical dotted lines show the adopted borders of the regions for the analysis of their integrated spectra in Fig.~\ref{fig:bpt}.}}
    \label{fig:ratios_dist}
\end{figure*}
}

\newcommand{\spectraexamples}{
\begin{figure*}
    \centering
    \includegraphics[width=\linewidth]{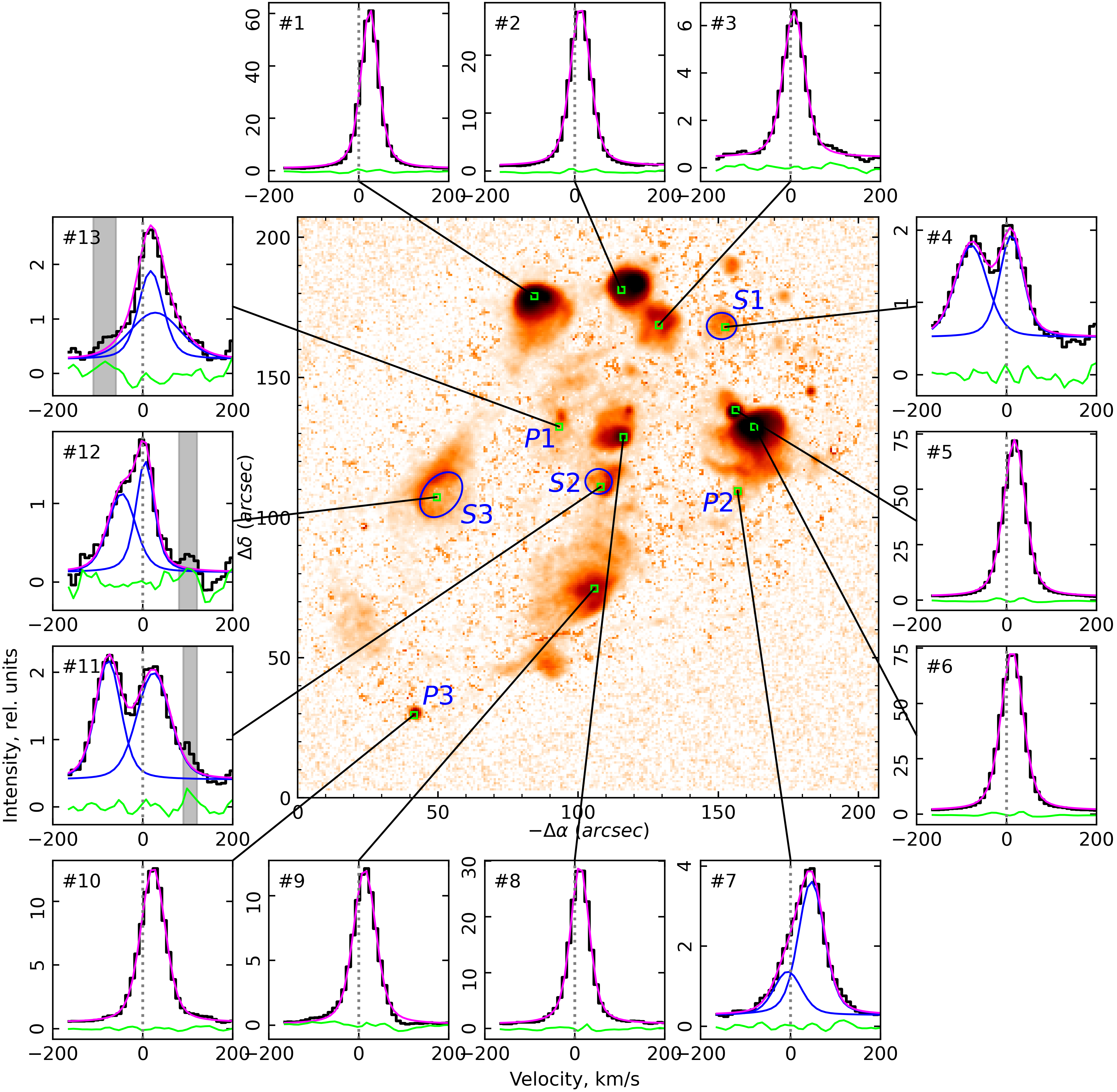}
    \caption{\Ha brightness distribution in the central region of Sextans B. The regions with elevated \Ha velocity dispersion discussed in the text are denoted as S1--S3 and P1--P3. Blue ellipses correspond to borders of the shells for the regions S1--S3. 
    The side panels demonstrate \Ha line profiles extracted in the apertures shown by light green rectangles on the \Ha map. The results of their best-fit modelling with a single- or two-component Voigt are shown by magenta lines, and the blue lines denote the individual components. The green colour shows residuals after subtraction of the best-fit model from the observed line profiles. \revone{The velocity scale corresponds to the residual velocity after subtraction of the galaxy rotation model. The grey stripes indicate residual artefacts from sky line subtraction.}}
    \label{fig:profs}
\end{figure*}
}

\newcommand{\bptdiagram}{
\begin{figure*}
    \centering
    \includegraphics[width=\linewidth]{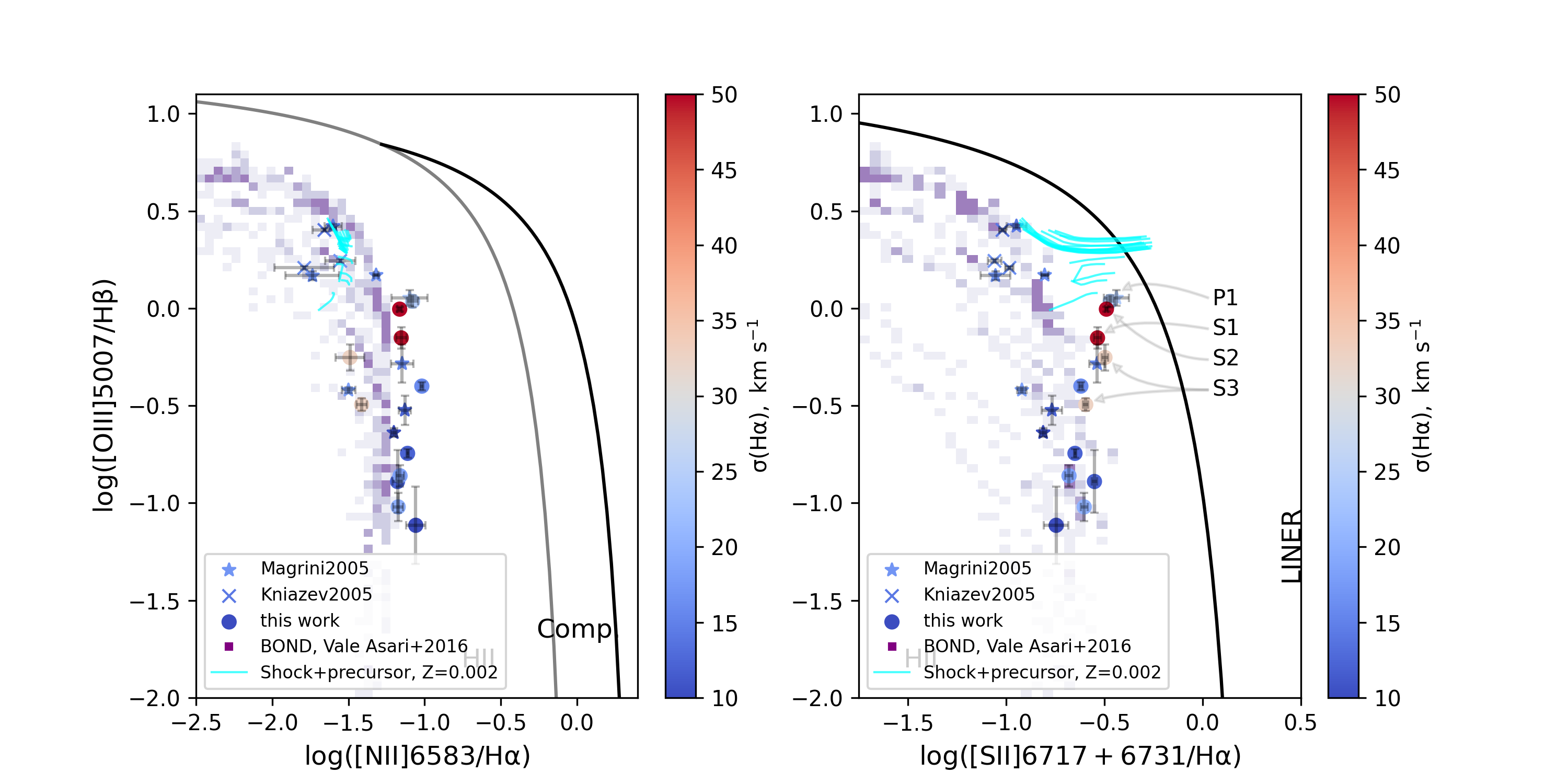}
    \caption{Diagnostic BPT \citep{BPT} diagrams for the nebulae with obtained spectral data. The left-hand panel shows \OIIIHb vs \NIIHa line ratios, and the right-hand panel demonstrates \OIIIHb vs \SIIHa diagnostic diagram. The measurements obtained from our data are shown by circles, while the crosses and asterisks show the measurements from the literature (\citealt{Kniazev2005} and \citealt{Magrini2005}, respectively). The symbols are colour-coded according to the measured \Ha velocity dispersion from the FPI data. The black curved line is the demarcation line between the regions with the dominated photoionisation and non-photoionisation mechanisms of gas excitation from \citet{Kewley2001}, and the grey curved line from \citet{Kauffmann2003} separate the regions with pure photoionisation and the composite mechanisms. These curves are valid for solar metallicity. Photoionisation models from \citet{Vale2016} for the $12 + \log(\mathrm{O/H}) = 8.0$ and $\mathrm{\log(N/O)} = -1.5$ are shown as a purple 2D histogram with their colour encoding the density of the model grid. Cyan lines show the shock+precursor model grid from \citet{Alarie2019} for metallicity $Z = 0.002$ ($\sim 0.15Z\odot$).}
    \label{fig:bpt}
\end{figure*}
}
\newcommand{\isigma}{
\begin{figure*}
    \centering
    \includegraphics[width=\linewidth]{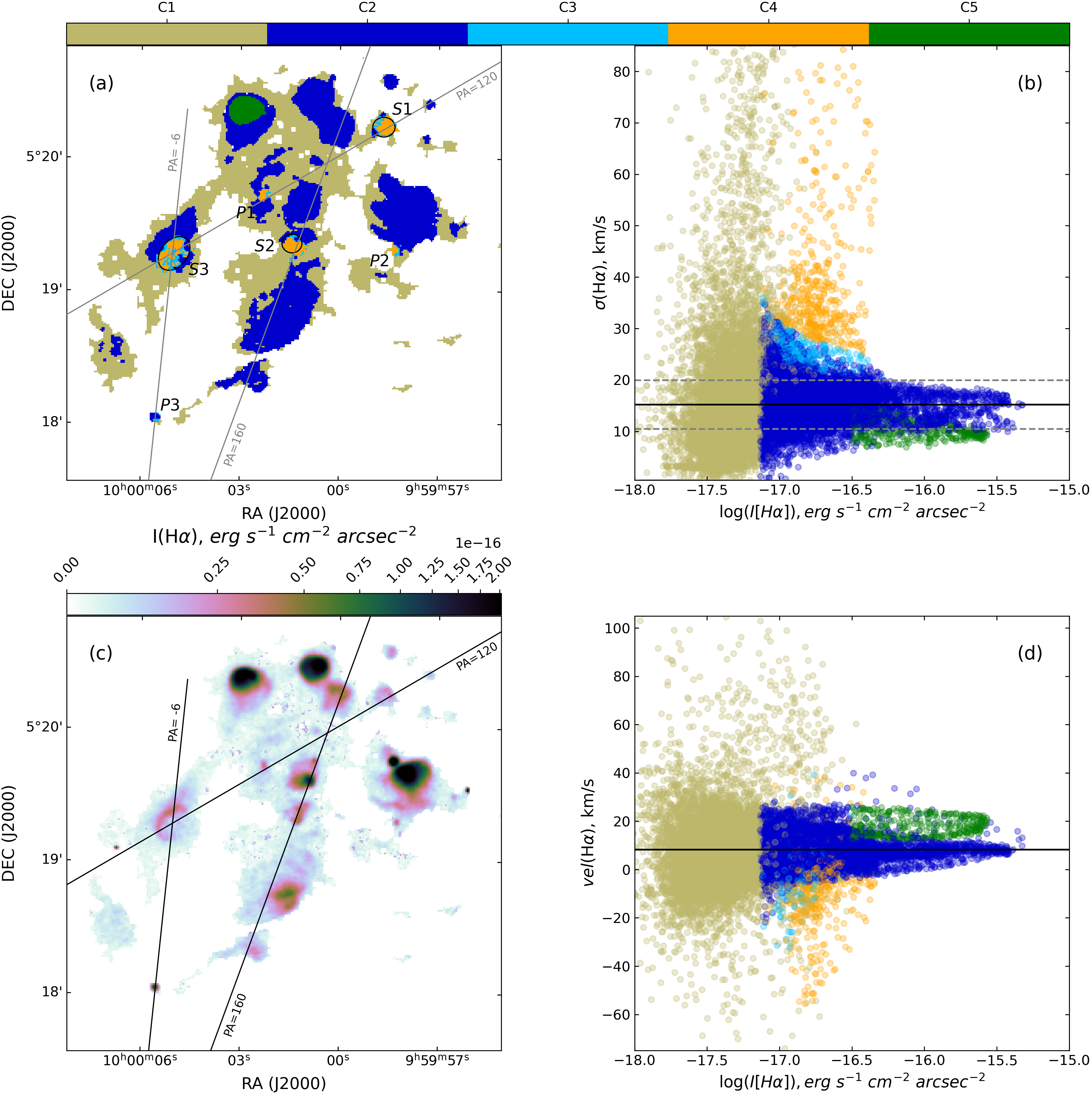}
    \caption{Classification of the pixels in FPI data based on the measured \Ha brightness and velocity dispersion. Panel (a) shows the classification map colour-coded according to the $\mathrm{I} - \sigma$ diagram (Panel b) according to the criteria described in the text. Defined classes correspond to the DIG component (C1), unperturbed \HII regions (C2), and the regions of elevated velocity dispersion, including expanding superbubbles (C3 and C4). Green colour (C5) mark the position of the brightest \HII region with high velocity residuals on the diagrams. Panel (c) gives the \Ha brightness map, for reference, in the same scale as the classification map. Panel (d) shows the diagram of the line-of-sight \Ha velocity (corrected for circular rotation of the galaxy) versus the logarithm of the \Ha brightness. Straight lines on the panels (a) and (c) show the localisation of the slits during the spectral observations, \revone{where PA = $-6^\circ$ corresponds to the TDS/CMO data, and other correspond to the SALT data.}}
    \label{fig:isigma}
\end{figure*}}

\title[Ionised gas in star-forming regions of Sextans~B]{Stellar feedback impact on the ionized gas kinematics in the dwarf galaxy Sextans~B.}

\author[I.~S.~Gerasimov et al.]{
Ivan~S.~Gerasimov,$^{1,2}$\thanks{E-mail: gerasimov.is18@physics.msu.ru (ISG)}
Oleg~.V.~Egorov,$^{3,1,2}$\thanks{E-mail: oleg.egorov@uni-heidelberg.de (OVE)}
Alexei~V.~Moiseev,$^{2,1}$
Alexei~Yu.~Kniazev,$^{4,5,1,2}$ \newauthor
Tatiana~A.~Lozinskaya,$^{1}$
Evgeniya~S.~Egorova$^{3,1}$
\\
$^{1}$ Lomonosov Moscow State University, Sternberg Astronomical Institute,
	Universitetsky pr. 13, Moscow 119234, Russia
\\
$^{2}$ Special Astrophysical Observatory, Russian Academy of Sciences, Nizhnii Arkhyz 369167, Russia
\\
$^{3}$ Astronomisches Rechen-Institut, Zentrum f\"{u}r Astronomie der Universit\"{a}t Heidelberg, M\"{o}nchhofstra\ss e 12-14, D-69120 Heidelberg, Germany
\\
$^{4}$South African Astronomical Observatory, PO Box 9, 7935 Observatory, Cape Town, South Africa\\
$^{5}$Southern African Large Telescope, Cape Town, 7935, South Africa
}

\date{Accepted XXX. Received YYY; in original form ZZZ}

\pubyear{2023}

\begin{document}
\label{firstpage}
\pagerange{\pageref{firstpage}--\pageref{lastpage}}
\maketitle

\begin{abstract}
We investigated the ionised and atomic gas kinematics and excitation state in the central region of ongoing star formation of the nearby low-metallicity dwarf galaxy Sextans B. The analysis is based on the new observations performed in \Ha emission line with high resolution ($R \sim 16000$) scanning Fabry-Perot interferometer at the 6-m BTA SAO RAS telescope, and on the long-slit spectral observations at the 9.2-m SALT and 2.5-m CMO SAI MSU telescopes. Strong non-circular gas motions detected in the studied regions probably resulted from the off-plane gas motions and impact of stellar feedback. We identified six regions of elevated \Ha velocity dispersion, five of which exhibit asymmetric or two-component \Ha line profiles. Three of these regions are young ($<1.1$~Myr) expanding ($V_\mathrm{exp} \sim 25-50 \kms$) superbubbles. We argue that at least three regions in the galaxy could be supernova remnants. We conclude that supernovae feedback is the dominant source of energy for superbubbles in Sextans~B, which is expected for such a low metallicity, although we cannot rule out a strong impact of pre-supernova feedback for one superbubble. 

\end{abstract}




\section{Introduction}

\interfootnotelinepenalty = 10000
\addtocounter{footnote}{-1}

The impact of stellar feedback on the interstellar medium (ISM) is one of the key topics in star and galaxy formation and evolution. Massive stars are generally considered the main sources of energy and momentum influencing the surrounding ISM of low-mass galaxies on both local and global scales. The ionising radiation of OB stars, as well as the mechanical energy of stellar winds and supernova explosions are heating gas, forming the cavities, shells, ordered outflows, and chaotic turbulent motions in the gaseous disc of a galaxy \citep{Krumholz2014, Klessen2016}. Energy injected into the ISM by massive stars is one of the main factors affecting gas turbulent motions in galaxies \revone{\citep[see e.g.][]{Tamburro2009, Moiseev2015, Egorov2023}}. The influence of multiple stellar winds and supernovae leads to the \revone{formation} of large superbubbles and shells with sizes up to \revone{the kiloparsec scale, which are} visible in many nearby galaxies of all masses \revone{\citep{Kim1998, Bagetakos2011, Egorov2017, Egorov2018, Pokhrel2020, Gerasimov2022, Watkins2023}}. 



A low-metallicity environment in nearby dwarf galaxies is of especial interest for studies of stellar feedback. Because of the weak gravitational potential, absence of density waves, and thick gas-rich disc in dwarf galaxies, massive stars can form large long-living shells. Even a single WR star might have \revone{a substantial effect on the evolution of the low-mass and metallicity systems} \cite[see, e.g.][]{Lozinskaya2003, Kehrig2016, Kehrig2018}. The lack of metals leads to lower energetics of pre-supernovae stellar feedback \citep{Vink2001, Bjorklund2021} resulting in the lower kinetic energy of the expanding bubbles \citep{Egorov2023}. Meanwhile, \revone{how the contribution of different stellar feedback mechanisms changes with metallicity is yet to be quantified} \citep[see][for some studies]{McLeod2019, McLeod2021, Ramachandran2019, Barnes2022, Gerasimov2022, Egorov2023}. 


In this work, we investigate the star-forming regions of the nearby low metallicity ($\mathrm{Z} \sim 0.13\mathrm{Z}_\odot$; \citealt{Kniazev2005, Magrini2005}) dwarf galaxy Sextans B. The galaxy is located at the outskirts of the Local Group in the small association with 3 other galaxies (Sextans A, NGC 3109 and Antlia dwarf; \citealt{Bergh1999, Tully2002}) isolated from massive galaxies. The main properties of Sextans B compiled from the literature are given in Table~\ref{tab:main_properties}.

\RGBHaFUVHI

\mainproperties


The star formation history in Sextans B exhibits a very strong star formation rate for previous 1-2 Gyrs \citep{Grebel1997, Grebel1999}, with the presence of young stars indicating a low-rate star formation in the recent epoch \citep{Tosi1991}.  
Recent star formation activity (on the scale of 100 Myr traced by far-ultraviolet; FUV) is concentrated in the central 2 kpc, while the extended \HI disc reveals multiple supershells \citep{Pokhrel2020} extending far beyond the central area (see Fig.~\ref{fig:map}). Some of \revone{these \HI supershells, which lack the young stars,} might be created by previous generations of massive stars \citep[e.g.][]{Weisz2009, Warren2011}, and the central \HI holes  are clearly associated with the FUV emission from the recent star formation activity. Current star-formation takes place in a dozen of HII regions \citep{Skillman1989, Strobel1991} residing in the rims of central \HI holes. Such morphology of ISM points to the propagation (or maybe even triggering) of star formation activity during the last several tens of Myr \citep[e.g.][]{Egorov2017, Mondal2019}.
The location of five planetary nebulae demonstrates low-efficiency star formation more distinct from the centre \citep{Magrini2005}. 

\footnotetext{created with the \textsc{multicolorfits} python package \citep{Cigan2019}}

In this paper, we present the first detailed study of ionised gas kinematics in star-forming regions of Sextans~B. We aimed to compare the global ionised and atomic gas kinematics in the star-forming complex and identify and analyse the regions of prominent supersonic ionised gas motions that could be associated to the strong impact of stellar feedback onto the ISM. Our analysis is based on the new observations performed with a scanning Fabry-Perot interferometer (FPI) in the $H\alpha$ line at the 6-m telescope of the Special Astrophysical Observatory of the Russian Academy of Sciences (SAO RAS), long-slit spectral observations with the 9.2-m South African Large Telescope (SALT) and 2.5-m telescope of the Caucasian Mountain Observatory of Sternberg Astronomical Institute \revone{of Moscow State University} (CMO SAI MSU), and archival data. 

This paper is organised as follows. Section~\ref{sec:observations} contains information on the observations obtained, the archival data used, and the data reduction procedures. Section~\ref{sec:analysis} describes the analysis of the ionised and neutral gas kinematics and the gas ionisation conditions in the star-forming regions of Sextans~B. In Section~\ref{sec:discussion}, we discuss the energetics and nature of the identified regions with strong non-circular motions in the galaxy. Section~\ref{sec:summary} summarizes our main conclusions.

\section{Observation and data reduction}
\label{sec:observations}

\observations

In this Section we describe new and archival data that where used in our study. Table ~\ref{tab:observations} contains information on our new observations performed at the 6-m telescope of SAO RAS, 
at 9.2-m Southern African Large Telescope \citep[SALT hereafter;][]{2006SPIE.6267E..0ZB,2006MNRAS.372..151O}, and the 2.5-m telescope CMO SAI MSU. 

\subsection{Scanning Fabry-Perot Interferometer observations} \label{obs:FPI}
The observations have been conducted in the prime focus of the SAO RAS 6-m telescope using a scanning Fabry-Perot interferometer (FPI) mounted in the multi-mode focal reducer \mbox{SCORPIO-2}~\citep{scorpio2}. 

The operating spectral range around the \Ha emission line was cut by a bandpass filter with $\mathrm{FWHM} \approx 14$~\AA.  During the observations, there were consecutively obtained 40 interferograms with different distances between the FPI plates with a single exposure 180~s. The data reduction was performed in the \textsc{idl} environment with the \revone{pipeline} described in detail by \citet{Moiseev2021}. After initial reduction, sky line subtraction, seeing and photometric corrections, made using reference stars, and wavelength calibration, these observations were combined in data cubes, where each pixel 
contains a 40-channel spectrum with spectral resolution about 0.48 \AA\ (22~$\kms$) and the \revone{separation between channels} of 0.22 \AA\ (10.1~$\kms$).

The analysis of emission line profiles was carried out with the software described in \citet{Moiseev2008} using single-component Voigt fitting, \revone{which yields a distribution of the relative \Ha emission line flux, line-of-sight velocity and velocity dispersion (corrected for instrumental broadening). Typical errors of the velocity dispersion measurements do not exceed $3 \kms$ for signal-to-noise ratio above 10 (see fig. 2 in \citealt{Moiseev2015}). } 
For the individual regions described in Section~\ref{sec:multi-component} we perform multi-component Voigt fitting to the line profile to derive the separation between the different kinematic components.

\subsection{Long-slit spectroscopic observations}
We obtained two long-slit spectra with different position angles (PA) at SALT
with the Robert Stobie Spectrograph \citep[RSS hereafter;][]{2003SPIE.4841.1463B, 2003SPIE.4841.1634K}.
The Volume Phase Holographic (VPH) grating, PG900, was used for observations with the PC0385 cutting filter covering the spectral range
around $4350-7390$\AA\ with a final reciprocal dispersion of about 0.96\AA\ per pixel. 
A slit width of 1\farcs5 resulted in a spectral resolution of 4.8~\AA\ (FWHM).
The slit positions were chosen to cross the regions with high velocity dispersion in \Ha line identified from the FPI observations (see Sec.~\ref{sec:multi-component}). 

The preliminary data reduction was performed with the standard SALT science pipeline, which includes gain correction, and bias and overscan subtraction.  Combining several expositions and long-slit data reduction was carried out with procedures described at \cite{Kniazev2008, Kniazev2022}.

In addition, we observed one long-slit spectrum with the Transient Double-beam Spectrograph at 2.5-m telescope CMO SAI MSU \citep{Potanin2020}. The observations were conducted as a short DDT program aiming to obtain the spectrum of one luminous blue variable (LBV) candidate (Sec.~\ref{sec:point_sources}) and one of the region with high velocity dispersion (Sec.~\ref{sec:multi-component}). Data reduction was carried out with the standard procedure described at \citet{Potanin2020}, which includes bias and dark subtraction, wavelength calibration, flat correction, and sky subtraction. The flux calibration \revone{was} performed using the spectrum of A0V HIP 107555 standard observed immediately after the object. We note that the data were taken in moderate and unstable weather conditions (high airmass, unstable seeing) thus limiting our analysis by only strongest emission lines.

We measured the emission line fluxes by fitting a single Gaussian profile to the integrated spectra of the regions (or individual pixels along the slits). \revone{The extracted spectra for the individual regions are provided in Appendix \ref{sec:appendix}}. The measured fluxes were corrected for reddening estimated from the Balmer decrement assuming a theoretical ratio of $\rm {\it F}(H\alpha)/{\it F}(H\beta) = 2.86$ for case B and electron temperature $T_e = 10\,000$~K typical for \HII\ regions \citep{Osterbrock2006}, and \cite{Cardelli1989} reddening law parameterized by \cite{Fitzpatrick1999}. We did not perform any modelling or subtraction of the underlying stellar absorption lines because of their negligible contribution to the emission spectra in Sextans~B.


\subsection{Archival data}

\subsubsection{HI 21 cm VLA data}

We used archival Very Large Array (VLA) \HI 21 cm data from the LITTLE THINGS survey \citep{Hunter2012} to compare the morphology and kinematics of ionised and atomic gas (Sec.~\ref{sec:velocities}). In this work, \revone{we analyse the natural-weighted (NA) data cube, which has} a velocity scale of $1.3 \kms$ per channel and angular resolution of $\mathrm{beam}_{\mathrm{NA}} = 17.9 \times 17.5$~arcsec. The published \revone{zero and first statistical moment maps} are used to trace the \HI 21 cm line flux and velocity, respectively.

\subsubsection{MUSE/VLT data}

We supplement our analysis with publicly available archival integral-field spectral observations with the Multi Unit Spectroscopic Explorer (MUSE; \citealt{Bacon2010}) at Very Large Telescope (VLT). MUSE observations covered a northern part ($1\arcmin\times1\arcmin$) of star-forming regions in Sextans~B with a spatial sampling of $0.2''$ per pixel. Data were obtained within programme 106.210Z.003 (PI: Bian) with a total exposure time of 4094s under seeing conditions $\sim 1.5$~arcsec. Here we use reduced data cube downloaded from the ESO Archive.


In this paper, we do not perform detailed analysis of the MUSE spectra, but only use it to validate our measurements of \Ha velocity with FPI. For that, we cut out wavelength range encompassing a window of $\pm 10$~\AA{} around \Ha line from the entire data cube and fitted a single-component Gaussian to the \Ha line profile. The obtained velocity maps are considered further in Sec.~\ref{sec:velocities}.




\section{Analysis}
\label{sec:analysis}

\subsection{Velocities of ionised and atomic gas}
\label{sec:velocities}
\velocitymap

In this Section we analyse line-of-sight velocities of the ionised and neutral ISM in the central star-forming regions of Sextans B. Fig.~\ref{fig:velocitymap}a shows \HI 21~cm first moment map, which is considered as atomic gas velocity field, while \revone{ionised gas velocity distribution shown in Fig.~\ref{fig:velocitymap}g was obtained by Voigt profile fitting as described in Sec.~\ref{obs:FPI}}. The map of \Ha line-of-sight velocities obtained from the MUSE data is shown on the inset panels in that Figure.

From the initial comparison, we noticed a large systematic offset (by $\sim 31 \kms$) between the velocities of the ionised gas (from FPI) that could only partially be explained by the physical differences between the kinematics of these two components of the ISM (e.g. by the warp of the \HI disc). 
Meanwhile, the \Ha velocities extracted from the MUSE data are in good agreement with the \HI velocities in the same $1 \times 1$ arcmin field. At the same time, the relative velocities in the FPI and MUSE data are very similar (Fig.~\ref{fig:velocitymap}g,h). We thus conclude that the absolute wavelength calibration of our FPI data \revone{for this galaxy} has a systematic offset. Therefore, we corrected all measured FPI line-of-sight velocities by $31 \kms$ to match their distribution from the MUSE data. 

The \Ha gas kinematics does not exhibit noticeable signatures of circular rotation, probably because all \Ha emission is concentrated in the central region of the galaxy. Indeed, the more extended \HI disc reveals a clear circular rotation pattern (Fig.~\ref{fig:velocitymap}a). Moreover, all \HII regions in the galaxy are located in the rims of the \HI supershells (see \HI morphology in Fig.~\ref{fig:velocitymap}d compared to the \Ha flux distribution in Fig.~\ref{fig:velocitymap}f), and thus their bulk kinematics could be significantly affected by the stellar feedback that created these shells. Finally, the local gas kinematics is regulated by stellar feedback from the sites of ongoing star formation. In order to isolate the influence of stellar feedback from the bulk gas motions due to the galaxy rotation, we model the velocity field and subtract it from the observational data following the `derotation' procedure described in \cite{Egorov2014}. 


The circular rotation model was recovered from the \HI 21 cm data and kinematic parameters derived by \cite{Oh2015} and \cite{Namumba2018}. \revone{For our analysis, we adopted the kinematic position angle, inclination, systemic velocity, and position of the rotation centre derived by \citet{Namumba2018} and listed in Table~\ref{tab:main_properties}. These authors}  found that the positional angle continuously changes from $\sim 47^\circ$ at the radii 50 arcsec to $\sim 60^\circ$ at the radii 350 arcsec, and it is almost constant on larger distances. We iteratively fitted the tilted ring model to the observed \HI first moment map as described in \citet{Moiseev2014}. In the first iteration, we fixed all parameters except the rotation velocity at a particular radial distance. After masking the regions with the residual velocities $> 10 \kms$ (as corresponding regions could be significantly affected by stellar feedback), we relaxed the position angle and constructed the model shown in Fig.~\ref{fig:velocitymap}(c). The resulting circular rotation model was subtracted from the \HI and \Ha velocity fields (Figs.~\ref{fig:velocitymap}e,h), and also from the FPI data cube for further analysis.

\spectraexamples

Both \HI 21~cm and \Ha residual velocity fields exhibit strong non-circular motions. 
As for \HI kinematics, its velocity field in the centre of \revone{the} galaxy is strongly affected by \HI holes, whose approaching components produce the blue-shifted residuals visible in Fig.~\ref{fig:velocitymap}e. The area with largest red-shifted \HI residuals is also associated with the expanding \HI supershell from the \citet{Pokhrel2020} list revealed in the \HI flux distribution (Fig.~\ref{fig:velocitymap}d). The outskirts of Sextans B reveal kinematic features visible on the residual map (Fig.~\ref{fig:velocitymap}b) that are typically observed in warps \citep{Kamphuis2015MNRAS.452.3139K,Wang2023ApJ...944..143W}. The warp of the \HI disc can also be partly responsible for the large residuals in Fig.~\ref{fig:velocitymap}e. 

Several clear outliers from the circular rotation are noticable in \Ha residual velocities (Fig.~\ref{fig:velocitymap}h). The brightest \HII region (H10 according to the list of \citealt{Strobel1991}) has the most prominent red-shifted velocity. The less obvious structures denoted as S1--S3 are visible both in the \Ha velocities (they exhibit blue-shifted residuals) and in the velocity dispersion maps (regions with clearly elevated $\sigma(\mathrm{H}\alpha$) in Fig.~\ref{fig:velocitymap}i; see Section~\ref{sec:multi-component} for details). Finally, large red-shifted residuals are also measured in the diffuse emission to the south of the S2 regions. The latter could be explained by uncertainties in the circular rotation models and differences in the angular resolution of the FPI data and the model (e.g., due to rapidly changing PA in the central part of the galaxy), but all the other features are also visible in the observed \revone{velocity} field in Fig.~\ref{fig:velocitymap}g. 

Therefore, from the described qualitative analysis of the \HI and \Ha velocity fields, we find strong non-circular gas motions in both components. The extended areas with blue-shifted residuals in \HI are probably caused by the expansion of the \HI supershells and the warp of the \HI disc of Sextans~B. We will explore the small-scale non-circular motions of the ionised gas \revone{in more detail below}.



\subsection{Identification of the regions with multi-component H$\alpha$ line profile}


\label{sec:multi-component}

From visual inspection of the \Ha map, we identified several medium-size ($\sim 100$~pc) shell-like structures in Sextans~B. \revone{Three of them} coincide with the regions S1--S3 exhibiting the blue-shifted residual velocities (Fig.~\ref{fig:velocitymap}h). 
\revone{We analysed the \Ha line profiles of these shell-like structures to decompose them onto the kinematically distinct components from the approaching and receding sides, or to uncover the underlying broad components. These signatures would confirm the ongoing expansion of the shells}. We show the line profiles extracted for candidates \revone{to expanding shells in the pixels exhibiting local maximum of dispersion velocity distribution} on Fig.~\ref{fig:profs}. \revone{The regions S1-S3} have asymmetrical \revone{line profiles that} could be decomposed into at least two components with velocity separation between them up to $100 \kms$, which is indicative of their expansion with velocities up to $50 \kms$.
\distproperties

\isigma
\spectrumpaotos
\revone{In previous works \citep[e.g.][]{Egorov2021, Gerasimov2022, Yarovova2023, Egorov2023}, the `$\mathrm{I}-\sigma$' diagram} (\Ha velocity dispersion against the logarithm of \Ha surface brightness for individual pixels) \revone{was shown to be an efficient tool for identifying the expanding bubbles} and the regions of supersonic gas motions in the ISM. This diagram was first proposed by \citet{Munoz1996} and further developed by \citet{Moiseev2012}. As was shown by these authors, the expanding superbubbles, high energy stellar sources (like Wolf-Rayet and LBV stars, SNRs), diffuse ionise gas (DIG) and normal \HII regions occupy different parts of the $\mathrm{I}-\sigma$ diagram, which \revone{simplifies their classification compared to the analysis of the velocity dispersion map alone. 
Following the same approach as in the papers mentioned above, we present the analysis of the $\mathrm{I}-\sigma$ diagram based on the FPI data for Sextans~B in Fig.~\ref{fig:isigma}.}

\revone{To classify different pixels on the $\mathrm{I}-\sigma$ diagram (shown in Fig.~\ref{fig:isigma}b)}, we followed the same approach as in \citet{Egorov2021}\footnote{Its most detailed description is given in \citealt{Egorov2023}, but in application to the MUSE data}.  
As a quick summary, we performed the following steps. First, we masked out all the pixels with the S/N < 3 in the \Ha line. On the resulting map (Fig.~\ref{fig:isigma}c) we masked out also regions where the number of adjacent pixels does not exceed 20, which we considered as probable ghosts or other artefacts. Then, we excluded pixels with \revone{\Ha brightness fainter than the median value for the galaxy} as dominated by DIG. This low surface brightness component is shown in olive colour in Fig.~\ref{fig:isigma} (denoted as C1). We measured the intensity-weighted average value of velocity dispersion ($\sigma_\mathrm{m}$) and standard deviation ($\sigma_\mathrm{std}$) for the rest of the pixels. The obtained value $\sigma_m = 15.1 \pm 4.8$ is considered typical for the \HII regions in Sextans~B and also consistent with what is normally measured in other dwarf galaxies \citep[e.g.][]{Moiseev+2015}. We consider all the pixels with $\sigma(\mathrm{H}\alpha) < \sigma_\mathrm{norm}$ as related to the unperturbed ionised ISM in \HII regions (denoted as C2; blue colour). Here $\sigma_\mathrm{norm}$ corresponds to the shape defined by Gaussian centred at $\sigma_m$ and with a standard deviation of $\sigma_\mathrm{std}$ (see details in \citealt{Egorov2023}). Cyan and gold colours (C3 and C4) on the $\mathrm{I}-\sigma$ diagram and classification maps correspond to pixels with the observed $\sigma(\mathrm{H}\alpha)$ below or above the Gaussian defined in the same way, but with standard deviation equal to $1.2\sigma_\mathrm{std}$. The physical meaning of these two regions is following: C4 are probably expanding superbubbles or the regions with high-velocity turbulent motions, and C3 is an intermediate class that could correspond to the pixels at the edges of superbubbles or turbulent regions with lower velocities than in C4-regions, but also sometimes could be artefacts. 
Finally, by green colour (C5) we encode the H10 region \revone{mentioned above in Sec.~\ref{sec:velocities} as exhibiting the highest} red-shifted non-circular velocities.  \revone{We note} that this region has the lowest $\sigma(\mathrm{H}\alpha)$ among all \HII regions of the same brightness and is a clear outlier from the $\mathrm{I}-\sigma$ diagram. 

\revone{The mentioned above regions S1, S2 and S3 with high blue-shifted velocities exhibit} the elevated velocity dispersion and are classified as C4 or C3 (Fig.~\ref{fig:isigma}d). In Sextans~B, we cannot clearly isolate the diagonal sequence on the $\mathrm{I}-\sigma$ diagram that is typically indicative of bright high-energy stellar sources \citep[e.g.][]{Moiseev2012, Yarovova2023}. There are, however, three relatively bright and compact sources with elevated $\sigma(\mathrm{H}\alpha)$ (classified as C3 and C4) -- they are denoted as P1--P3 and will be considered below. 

In Fig.~\ref{fig:isigma}d we show the line-of-sight velocity (after subtracting the circular rotation model) versus \Ha intensity diagram for the same pixels as on panel (b). It is evident from this plot that almost all pixels with elevated velocity dispersion classified as C4 exhibit also large deviations from the circular rotation model. Bi-modality of both diagrams on panels (b), (d) caused mostly by the brightest \HII region H10. The difference in $\sigma(\mathrm{H}\alpha$) of this region from the other \HII regions could be in principle explained by the differences in physical conditions, such as external gas pressure. The region H10 is indeed located towards the dense \HI cloud, while most of the other regions are observed in the more rarefied ISM. Two other bright \HII regions located in the relatively dense \HI gas (Fig.~\ref{fig:velocitymap}d) also exhibit relatively narrow line profiles (Fig.~\ref{fig:profs}, see profiles~\#2 and \#6). The difference in the bulk velocity of H10 relative to other \HII regions is less clear but could be produced by local non-circular off-plane motions not accounted in the constructed rotation model.



Summing up, from the analysis of the $\mathrm{I}-\sigma$, we found \revone{6 regions with elevated velocity dispersion}. Three of them (S1--S3) have shell-like morphology and two-component \Ha line profiles (\#4, 11 and 12 in Fig.~\ref{fig:profs}), which is a clear indication that they are expanding superbubbles of ionised gas. The other three regions (P1--P3) are relatively bright and compact (unresolved) regions. One of them (P3) has broadened \Ha line profile (\#10 in Fig.~\ref{fig:profs}), while two other regions exhibit asymmetric profiles that can be decomposed by two components (\#7 and 13 in Fig.~\ref{fig:profs}). In particular, a broad underlying component is detected in region P1. We will consider the possible source of the ionisation and the nature of all these regions in the next sections. Other line profiles in Fig.~\ref{fig:profs} not mentioned here correspond to the bright \HII regions and are given for comparison. 
\revone{In Table~\ref{tab:distproperties} we provide the main properties of line profiles shown in Fig.~\ref{fig:profs}.}

\bptdiagram

\subsection{Gas excitation state}
\label{sec:bpt}

To define the probable source of the regions with high-velocity non-circular motions of ionised gas, we analyse here their excitation state using the obtained long-slit spectral data. Several strong emission line ratios served as good diagnostics of the ionisation condition \citep[see e.g.][for a review]{Kewley2019} such as \OIIIHb\ (probe of the hardness of the ionising radiation), \SIIHa\ and \NIIHa\ (both are sensitive to shocks, but the latter is very dependent on the gas-phase metallicity). 
 Fig.~\ref{fig:ratios_dist} shows the distribution of the fluxes and ratios of several strong lines along the slits for the SALT spectra. We see an increase of \SIIHa from $\sim 0.2$ in the bright HII regions with low velocity dispersion up to $\sim 0.4$ in the regions S1 and S2 with elevated gas velocity dispersion. The latter value usually serves as the threshold for the SNRs at solar metallicity \citep{Dodorico1980, Long2018}. For the low surface brightness regions, we measured \SIIHa higher than 0.4, which is typical for DIG \citep{Haffner2009, Belfiore2022}.  The ratio of \OIIIHb\ does not exceed 2 in all studied regions, which indicates relatively low hardness of the ionising radiation for them.

In Fig.~\ref{fig:bpt}, we show the classical diagnostic BPT diagrams \citep{BPT, Veilleux1987} for the studied regions in Sextans~B \revone{(their spectra are provided in Appendix~\ref{sec:appendix})}. Our measurements from all three spectra are shown by circles colored according to the measured \Ha velocity dispersion. In addition, we show line ratio measurements for the \HII regions taken from \citet{Kniazev2005, Magrini2005} as the reference (shown by crosses and asterisks, respectively). We also overlaid \citet{Alarie2019} shock models for $Z\sim 0.15 Z_\odot$ (cyan lines) and 
photoionization models from \citet{Vale2016} for the same metallicity (purple histogram) for fixed values of metallicity $12 + \log\mathrm{(O/H)} = 8.0$ and $\rm \log(N/O) = -1.5$ (consistent with \revone{the measurements for} the \HII regions and PNe in Sextans~B, see \citealt{Kniazev2005, Magrini2005}). We choose the latter grid of photoionisation models among other available ones because it covers the range of ionisation parameters sufficient to cover our measurements for regions with low \OIIIHb ratio. It is shown in Fig.~\ref{fig:bpt} as 2D histogram with intensity coding the density of model grid in particular range of line ratios. Both photoionisation and shock grids were obtained from the Mexican Million Models database (3MdB) \citep{Morisset2015}. The measurements of line fluxes for the \HII regions from \citet{Kniazev2005, Magrini2005} are in good agreement with the photoionisation models that proves its validity for the comparison with our measurements for the regions with high velocity dispersion. 

In contrast to the high metallicity ISM \citep[e.g.][]{Allen2008}, the photoionisation and shock model grids partly overlap on BPT (i.e., both shocks and photoionisation could produce similarly low \SIIHa or \NIIHa line ratios) at such a low metallicity as for Sextans~B, which makes it difficult to disentangle between the two mechanisms. The strong dependence of \NIIHa on the metallicity makes this even more challenging. For example, in Fig.~\ref{fig:bpt} the \citet{Alarie2019} shock models cover the same range as photoionisation models from \citet{Vale2016} and match the line flux ratios measured for \HII regions by \citet{Kniazev2005, Magrini2005}. This could be because of insufficient precision of the \NIIHa predictions in the low-metallicity shock models \citep[see also similar comparison in][]{Egorov2021, Gerasimov2022}. In contrast, \SIIHa is much less dependent on metallicity and provides a more reliable instrument for shock diagnostics. Considering the velocity dispersion as a third parameter on such plots makes conclusions on the dominant excitation mechanism more reliable \citep[e.g.][]{Oparin2018, Law2021}.

We find that the measured line flux ratios for the regions with low velocity dispersion agree well with the predictions from the photoinisation models. At the same time, the regions with elevated velocity dispersion S1 and S2 slightly decline from the photoionisation models on the right-hand plot of Fig.~\ref{fig:bpt} and reside in the region that could be described by shock models. This is also true for the compact unresolved source P1, which has relatively low, but still elevated velocity dispersion compared to the other \HII regions, and also reveal broad underlying component (profile \#13 in Fig.~\ref{fig:profs}). Therefore, \revone{taking into account the high velocity dispersion} we expect that shocks contribute significantly to the excitation of these three regions. The region S3, which also has elevated velocity dispersion, shows significantly lower \OIIIHb and is in better agreement with the photoionisation models. Thus, the contribution of shocks to the overall ionisation balance is less important than the contribution of the photoionisation for this particular region. 



\section{Discussion}
\label{sec:discussion}

\subsection{Nature and energetics of the expanding bubbles}
\label{sec:shells_nature}
\bubbleparams


The analysis presented in Sections~\ref{sec:velocities} and \ref{sec:multi-component} revealed 6 regions with elevated velocity dispersion, while only three have sizes large enough to resolve them with our FPI observations. These three regions (S1, S2, S3) have shell-like morphology in \Ha and thus are probably expanding superbubbles. Here we measure their properties and estimate mechanical energy input from massive stars required for their formations. The results are summarised in Table~\ref{tab:bubble_params}, which contains coordinates of the centres of each superbubble, measured radii along the major and minor axes ($R_a$ and $R_b$, respectively, in arcseconds), their positional angle (PA), effective radius ($R_\mathrm{eff} = \sqrt{R_aR_b}/2$, converted to parsecs), kinematic age ($t_{\rm kin}$), expansion velocity ($V_{\rm exp}$), estimated mechanical luminosity ($L_{\rm mech}$) required for their formation, total kinetic energy $E_{\rm kin}$ of the bubbles, volumetric atomic number density ($n_\mathrm{0}$), and the equivalent number of O5V stars (N) necessary to form the wind-driven bubble with the derived properties.

The sizes of the bubbles measured from the \Ha map are 50-100~pc. From the decomposition of the \Ha line profiles onto two distinct kinematic components in the central parts of the bubbles, we can estimate their expansion velocities. We measured separation between blue- and red-shifted components assuming they are originating from the approaching and receding sides of the bubble, respectively. The corresponding expansion velocity $V_{\rm exp}$ is estimated as half of the separation between the components.

The evolution of a bubble/superbubble under the influence of the winds and supernova explosions can be described by classical models by \citet{Weaver1977, MacLow1988}. According to them, we estimate the kinematic age of the bubbles as 

\be t_{\rm kin} \simeq 0.6R_\mathrm{eff}/V_{\rm exp} \ee

According to the equations from \citet{MacLow1988}:

\be R_\mathrm{s}(t) \simeq 66(\frac{L_{38}}{n_0})^{1/5}t_\mathrm{kin}^{3/5}; \ee
\be V_\mathrm{exp}(t) \simeq 38.6(\frac{L_{38}}{n_0})^{1/5}t_\mathrm{kin}^{-2/5}, \ee

where $R_\mathrm{s}$ -- radius of the bubble in parsecs ($R_\mathrm{eff}$ in our case), $L_{38} = L_\mathrm{mech} \times 10^{-38}$ -- mechanical luminosity of central star cluster in $\rm erg\ s^{-1}$, and $t_\mathrm{kin}$ is expressed in Myr. Thus, we can estimate the $L_\mathrm{mech}$ required to produce the observed bubbles assuming they evolve in the homogeneous medium following this classical model:

\be L_\mathrm{mech} \simeq 3.99\times 10^{29}{{n_0}R^2 V_\mathrm{exp}^{3}}\ \mathrm{erg\ s}^{-1}, \ee
We estimate the gas density $n_0$ from the \HI observations. The column density $N_\mathrm{HI}$ is measured directly from the mean brightness in \HI line in the aperture around the S1--S3 objects following the conversion from \citet{Hunter2012}. From this we can derive $n_0 \simeq \mu n_\mathrm{HI}$, where $\mu = 1.4$ to take into account the contribution of He, and the \HI volume density averaged along the line-of-sight ($n_\mathrm{HI}$) can be derived from:

\be N_\mathrm{HI} = \int^{+\infty}_{-\infty}n_\mathrm{HI}\exp{\frac{-z^2}{2h^2}}dz = \sqrt{2\pi}hn_\mathrm{HI} \ee

The measurements of a scale height of gas disc ($h$) are very uncertain for Sextans B. \cite{Hunter2004, Stilp2013, Pokhrel2020} provided values $h = 480, 639$, and $280$~pc, respectively. In our work we adopted the largest value to obtain the lower limit for the volume density and required energy input to produce the bubbles. Finally, we corrected the $n_\mathrm{HI}$ for inclination $i=49^\circ$ (see Tab.~\ref{tab:main_properties}) and obtained


\begin{equation}
n_{0} = \frac{\mu N_\mathrm{HI} \cos{i} }{\sqrt{2\pi}h}.
\end{equation}

According to these estimates, all three regions S1--S3 are relatively young (0.4--1.2 Myr). The mechanical luminosity necessary for their formation (assuming constant energy injection) is $L_\mathrm{mech} \sim 4-10\times10^{36}$~erg~s$^{-1}$ that corresponds to the total energy $E_\mathrm{kin} \sim 0.1\times10^{51}$ erg injected into the bubble during their lifetime, which corresponds to their kinetic energy assuming the 100\% coupling efficiency. This value is in agreement with \revone{the measurements} for superbubbles in several other low-metallicity galaxies \citep[e.g.][]{Egorov2021, Gerasimov2022}. The estimated energy is much lower than the energy of the core collapse SN explosion ($10^{51}$~erg). However, the calculation above are based on the classical \citet{Weaver1977} model, which does not consider the inhomogeneities and turbulent nature of the ISM, as well as the radiative loss during the evolution of the bubbles \citep[see, e.g.][and references therein]{Krause2014, Lancaster2021}. Given that the typical value of the coupling energy efficiency measured for the simulated \citep[e.g.][]{Yadav2017} and observed \citep{Egorov2023} superbubbles is $\eta \sim 0.1-0.2$, we obtain the real total mechanical energy injected into the superbubbles S1--S3 during their lifetime
$E_{\rm mech} \sim E_{\rm kin} / \eta \sim 0.5-1\times10^{51}$. Thus, the energy of a single supernova explosion is sufficient to create these superbubbles. 

From the analysis of the spectra of these three superbubbles (Sec.~\ref{sec:bpt}) we conclude that shocks \revone{probably provide significant contribution} to the gas ionisation in S1 and S2. Together with the estimates of the required mechanical energy, this allows us to \revone{suggest} that these two regions are probable SNRs. The age of these bubbles is much lower than the typical lifetime of the massive stars with $M>8M_\odot$. This discrepancy could be explained if we assume that the currently observed kinematics was onset at the moment of SN explosion, whereas the pre-SN feedback from the progenitor star only precleared the ISM. Probably, before the SN explosion we would also observe the expanding bubble, but with a lower expansion velocity not measurable in our data. This hypothesis is consistent with the finding by \citet{Egorov2023} for the large sample of the superbubbles in the high-metallicity ISM of the nearby large star-forming galaxies: the measured kinematic age is in much better agreement with the time passed since the first SN explosion than with the total age of the central clusters.

We cannot be confident that the region S3 is also SNR. We measured significant expansion velocity $\sim 26 \kms$ for this region, however, we do not observe signatures of shocks in its spectrum (see Sec.~\ref{sec:bpt}). Furthermore, twice lower mechanical luminosity input is necessary to explain its formation. Therefore, we check here whether this (or any other) bubble could be formed by pre-SN feedback alone. 
For that, we estimate the number of O5I stars necessary to produce required mechanical luminosity for blowing the bubbles assuming that their expansion is driven mostly by stellar wind. We assume that the stellar model from \cite{Smith2002} for the star OB\#14 with $\mathrm{Z} = 0.2 \mathrm{Z_\odot}$ could describe the O5I stars in Sextans~B. This model yields mass-loss rate $\log(\dot{M}) = -5.48$ and wind velocity $v_{\inf} = 1860 \kms$. According to these values, a single O5I star at this metallicity should have mechanical luminosity $L_\mathrm{mech}(\mathrm{O5I}) = 0.5 \dot{M} v_{\inf}^2 = 3.62 \times 10 ^ {36}\ \mathrm{erg\ s}^{-1}$. Thus, a single O5I star could blow-out the S3 bubble. Several such stars are required for the other two regions. Given the relatively low intensity of recent star-formation in all these regions (traced by FUV, see Fig.~\ref{fig:map}), we do not expect to see many supergiant O stars in any of these regions, but a single such star in principle could be present in the region S3. We note that the necessary amount of O5V star is about 13 times larger (model OB\#3 from \citealt{Smith2002} provides values
$\log(\dot{M}) = -6.8$ and $v_{\inf} = 2330 \kms$). The estimated number of O5V stars is given in Table~\ref{tab:bubble_params}.

Thus, we conclude that the superbubbles S1 and S2 are probable SNRs, while the S3 (associated with the faint extended \HII region) can be a product of pre-SN feedback from a single massive supergiant star, or several O-type main sequence stars. We cannot exclude that it is SNR as well, although in that case photoinisation from the \HII region is dominating in the spectrum making the spectral signatures of shocks almost invisible (while they are still detected in the gas kinematics).




\subsection{Point like sources with high velocity dispersion}
\label{sec:point_sources}

In previous studies, \citet{Kniazev2005} and \citet{Magrini2005} analysed the spectra of 5 PNe found in Sextans~B. In our \Ha data, we identified several other point-like sources not coinciding with these PNe. Among them, 3 stellar-like sources (P1--P3 in Fig.~\ref{fig:profs}) have high \Ha velocity dispersion, and line profiles of two of them can be decomposed by two kinematically distinct components (profiles \#13 and 7 for objects P1 and P2, respectively). One of these objects (P1) even exhibits a broad underlying component with the \Ha velocity dispersion $\sim 51 \kms$. Such complex kinematics is indicative of an unresolved expanding bubble or an outflowing ionised gas in these regions. In particular, such objects could be nebulae around WR or LBV stars. Identification and analysis of such objects in low-metallicity ISM is especially important because they are very rare and the theoretical predictions of their properties are controversial \citep[see discussion in][]{Yarovova2023}. 

The position of the region P1 is similar to that of the S2 on the BPT diagram (see Fig.~\ref{fig:bpt}), so we assume that the shocks are significant there and probably responsible for the formation of the broad underlying component. The origin of these shocks is unclear. It could be due to a recent SN explosion, so that the resulting SNR is young and has not developed yet the prominent bubble that could be resolved in our data. Another scenario is that this region resulted from the outflows from two 
bright clusters visible in FUV (Fig.~\ref{fig:map}). We note that this object resides in the center of the \HI supershell (Fig.~\ref{fig:velocitymap}d) and relatively isolated from the nearest \HII regions located in the rims of the \HI supershells that makes first scenario more reliable. 

Another unresolved region with a 2-component \Ha line profile is P2. For this region, we do not have spectral observations, and thus we cannot judge its dominant excitation mechanism. We note that the second kinematic component is not necessarily physically associated with this region and could be the result of the overlap between two nebulae with different kinematics \citep[see Region \#24 in ][as example]{Egorov2021}, so additional spectral observations are essential to understand its nature.

Object P3 was reported by \citet{Masseylines} for the first time as an LBV candidate \revone{with the apparent magnitude $m_V = \mathrm{21.68^m}$ ($M_V = \mathrm{-4.1^m}$)}. This object is located on the south-eastern periphery of the central star-forming region, in the rims of the \HI hole in isolation from all other \HII regions. According to the FUV images, there was only weak recent star formation activity (see Fig.~\ref{fig:map}). However, the localisation of the object in the \HI rim (in contrast to the P1) allows us to assume that this star was formed as a result of star formation there. The LBV candidate \revone{exhibits} a slightly broadened single-component \Ha line profile ($\rm \sigma(H\alpha) \sim 22 \kms$, while $\rm \sigma(H\alpha) \sim 15 \kms$ is measured for \HII regions in the galaxy, see Sec.~\ref{sec:multi-component}). Observation of this \revone{object} is very difficult to perform because of its proximity to the bright foreground star. We obtained spectrum of this region with the 2.5-m telescope under bad seeing conditions and were able to measure only a few bright emission lines at  $>3\sigma$ level: [O~\textsc{ii}]3727+3729\AA, \Ha and H$\beta$ \revone{(see Appendix~\ref{sec:appendix})}. The presence of bright [O~\textsc{ii}] and non-detection of the \OIII\, emission are indicative of a low ionization parameter in the region\footnote{\revone{We cannot completely rule out the effect of atmospheric dispersion differently affecting the blue and green parts of the spectra. However, in case of its significant impact, we would expect the underestimation of [O~\textsc{ii}] rather than of [O~\textsc{iii}] line flux because the acquisition was performed in a red-band filter.}}. \revone{Given this, P3 is unlikely a nebula around an LBV star. However, }
it is not clear what the nature is of this region and what is responsible for the broadened and bright \Ha emission from this star. 
Overall, P3 is a very interesting object for further investigation. In particular, because of its proximity to a close bright star, it is possible to observe this nebula with adaptive optics at high angular resolution (about $0.03\arcsec$) with MUSE/VLT. Such observations would provide the resolved structure of the nebulae around the isolated low-metallicity hot star.







\section{Summary}
\label{sec:summary}

We investigated the ionised and neutral ISM in the central region of ongoing star formation of the nearby low-metallicity dwarf galaxy Sextans~B. We focused on the ionised gas kinematics, its relation to the neutral gas morphology and kinematics and gas excitation state. Our analysis relies on the new high spectral resolution observations ($R \sim 16000$) in \Ha line with a scanning Fabry-Perot interferometer at 6-m BTA SAO telescope, and on the long-slit spectral observations at the 9.2-m SALT and 2.5-m CMO SAI MSU telescopes. Our main findings from the analysis of these data are the following:

\begin{itemize}
    \item Current star-formation in the galaxy proceeds predominantly in the rims of \HI supershells probably created by previous generations of massive stars. Strong non-circular motions of ionised gas in some of the regions are indicative of off-plane gas motions.
    \item Six regions in the galaxy have elevated \Ha velocity dispersion compared to normal \HII regions. Most of them exhibit two-component \Ha line profiles. We argue that three spatially resolved objects are young (0.5--1.2 Myr) expanding superbubbles.
    \item We argue that at least three of the nebulae with high \Ha velocity dispersion are supernova remnants or resulted from the shocks impacting the ISM. 
    \item One of the unresolved sources with elevated \Ha velocity dispersion is an LBV candidate according to the literature \citep{Masseylines}. We detected only a few strong emission lines in its spectrum ([O~\textsc{ii}], \Hb and H$\alpha$) and thus cannot confirm its nature, although we did not find signatures of hard ionising radiation from this source. \revone{As a result, the origin of elevated velocity dispersion towards this region is still under question.}  
\end{itemize}


Overall, we conclude that combination of global perturbations of the gas disc and the influence of the stellar feedback is responsible for the observed peculiarities in the ionised gas kinematics of Sextans~B. The measured energy of the identified expanding superbubbles is consistent with the previous estimates made for the nebulae in several other low-metallicity dwarf galaxy \citep[e.g.][]{Egorov2021, Gerasimov2022} and lower than is typically measured in high-mass galaxies \citep{Egorov2023}. Together with those observations, our results indicate that typical energetics of superbubbles decrease at the low-metallicities. In the case of Sextans~B, supernovae are probable sources of energy for at least two of three superbubbles, and for at least one unresolved source. We cannot rule out the significant impact of pre-SN feedback in at least one of the studied bubbles, despite it is not expected to be strong in a low metallicity environment. 

\section*{Acknowledgements}
We are grateful to D.~Oparin and S.~Zheltoukhov for their help with the BTA/FPI and CMO/TDS observations, respectively, A.~Dodin for primary data reduction of CMO/TDS observations, and A.~Yarovova, K.~Vasiliev \revone{and O.~Maryeva} for the useful discussions during the preparation of the paper. \revone{We thank the anonymous referee for their useful comments, which improved the quality of our paper.}

This study is based on the data obtained at the unique
scientific facility the Big Telescope Alt-azimuthal SAO RAS and was supported under the Ministry of Science and Higher Education of the Russian Federation grant 075-15-2022-262 (13.MNPMU.21.0003).
Most of the long-slit spectral observations reported in this paper were obtained with observational program 2019-1-SCI-007 (PI: Kniazev) at the Southern African Large Telescope (SALT).
We are grateful to CMO SAI MSU staff for providing DDT time for the additional spectral observations. 
AK acknowledges support by the National Research Foundation (NRF) of South Africa.
This research made use of Astropy (http://www.astropy.org) a
community-developed core Python package for Astronomy
(Astropy Collaboration et al. 2013, 2018) and Astroalign
(Beroiz et al. 2020). We acknowledge the usage of the HyperLeda database (http://leda.univ-lyon1.fr; Makarov et al. 2014).

\section*{Data Availability}
The data underlying this article will be shared on reasonable request to the corresponding author. The reduced FPI data will be available in SIGMA-FPI data base\footnote{\url{http://sigma.sai.msu.ru}} (Egorov et al., in preparation).



\bibliographystyle{mnras}
\bibliography{SexB} 


\appendix

\section{Extracted spectra of the nebulae exhibiting asymmetrical \Ha line profile}
\label{sec:appendix}

\begin{figure*}
    \centering
    \includegraphics[width=\textwidth]{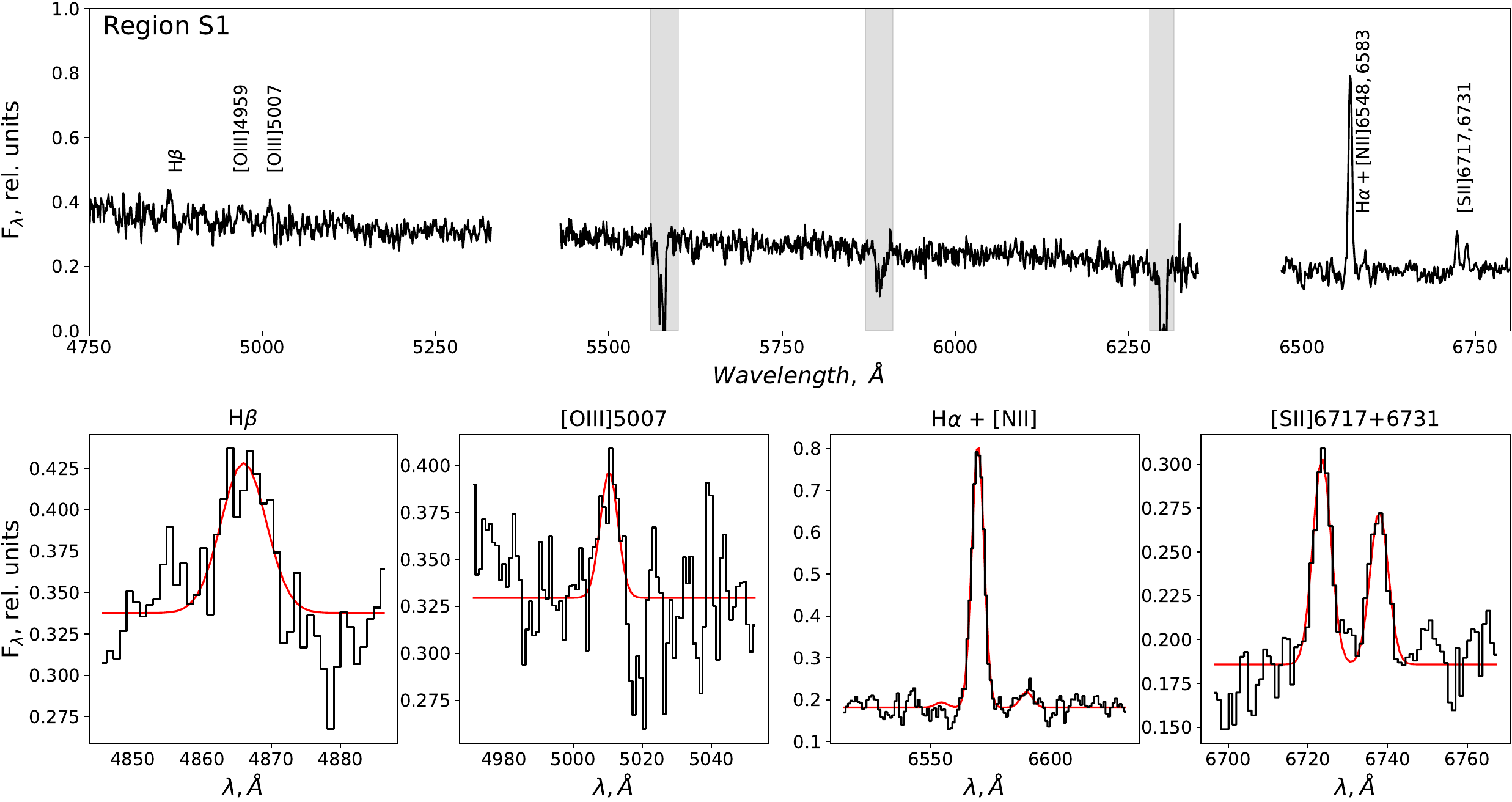}
    \caption{\revone{Top panel: Optical spectrum for the S1 region obtained with RSS at SALT. Vertical grey stripes denote regions with bright sky line emission producing residual artefacts after their subtraction. Bottom panels: Zoom-in on the brightest emission lines. The observed spectrum and the best-fit Gaussian model are shown by black and red lines, respectively.}}
    \label{fig:S1_spectrum}
\end{figure*}

\begin{figure*}
    \centering
    \includegraphics[width=\linewidth]{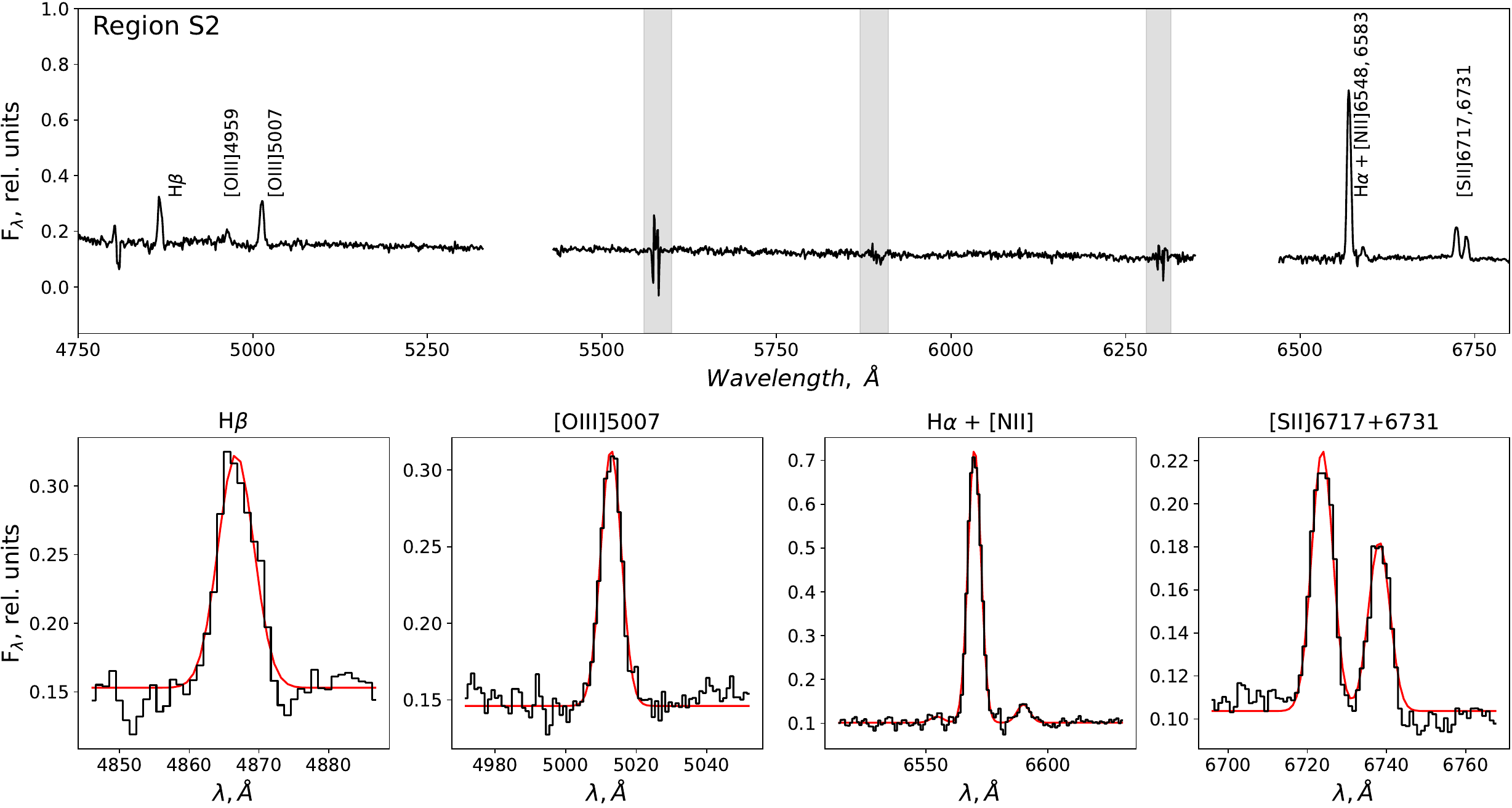}
    \caption{\revone{Same as in Fig.~\ref{fig:S1_spectrum} but for the S2 region.}}
    \label{fig:S2_spectrum}
\end{figure*}

\begin{figure*}
    \centering
    \includegraphics[width=\linewidth]{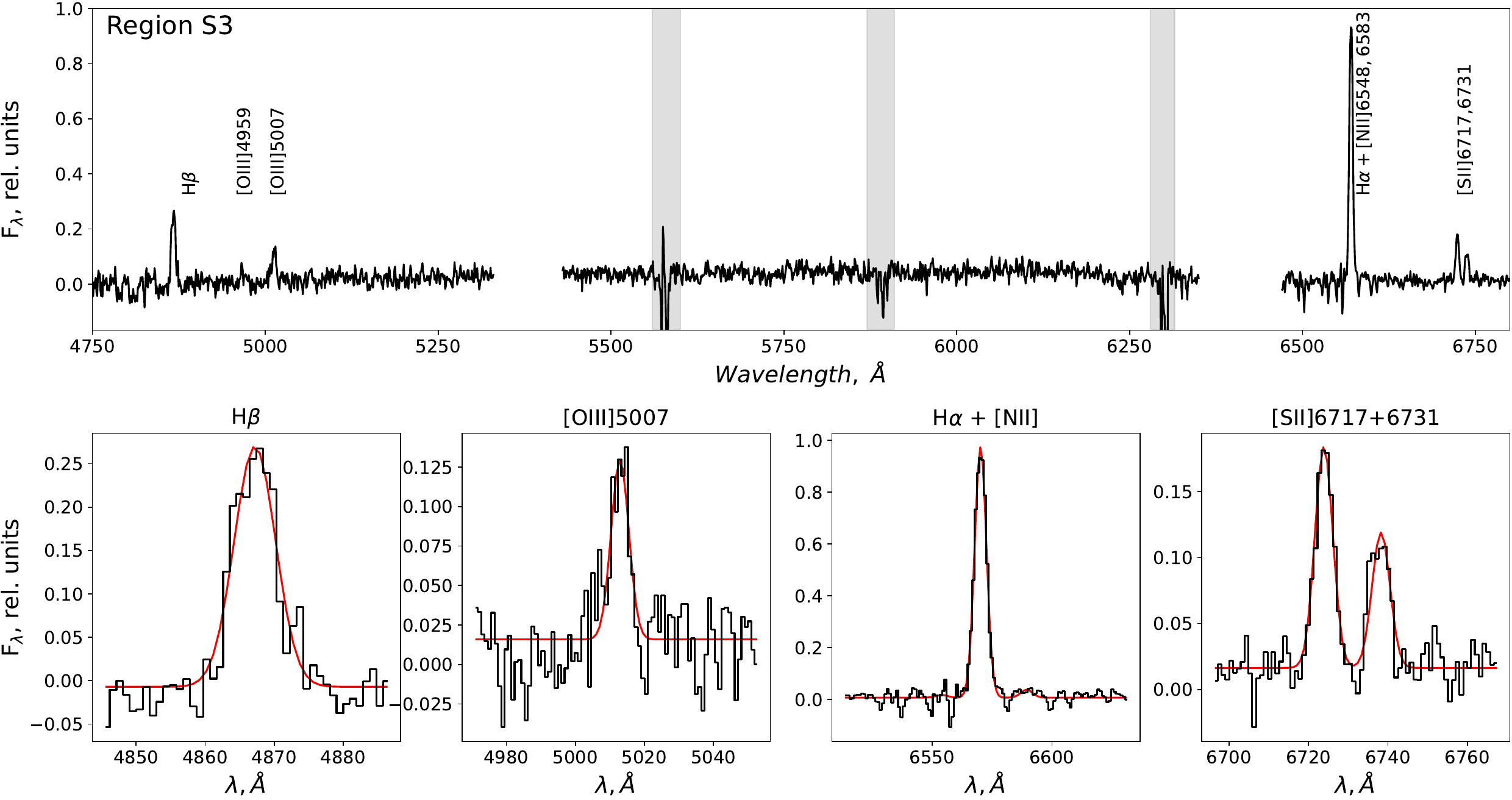}
    \caption{\revone{Same as in Fig.~\ref{fig:S1_spectrum} but for the S3 region.}}
    \label{fig:S3_spectrum}
\end{figure*}

\begin{figure*}
    \centering
    \includegraphics[width=\linewidth]{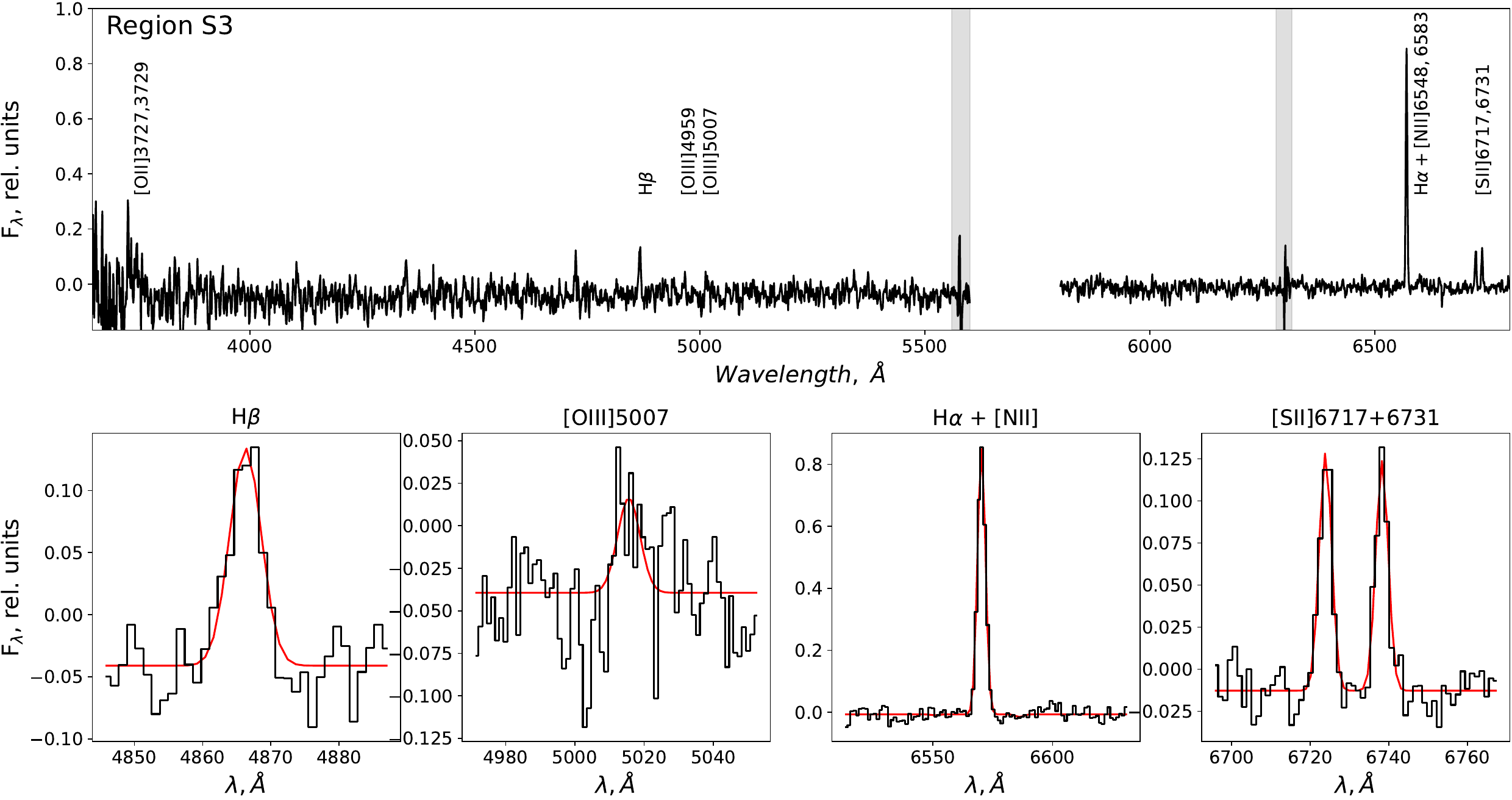}
    \caption{\revone{Same as in Fig.~\ref{fig:S3_spectrum} but for the spectrum obtained with TDS at CMO SAI MSU.}}
    \label{fig:S3_spectrum_CMO}
\end{figure*}

\begin{figure*}
    \centering
    \includegraphics[width=\linewidth]{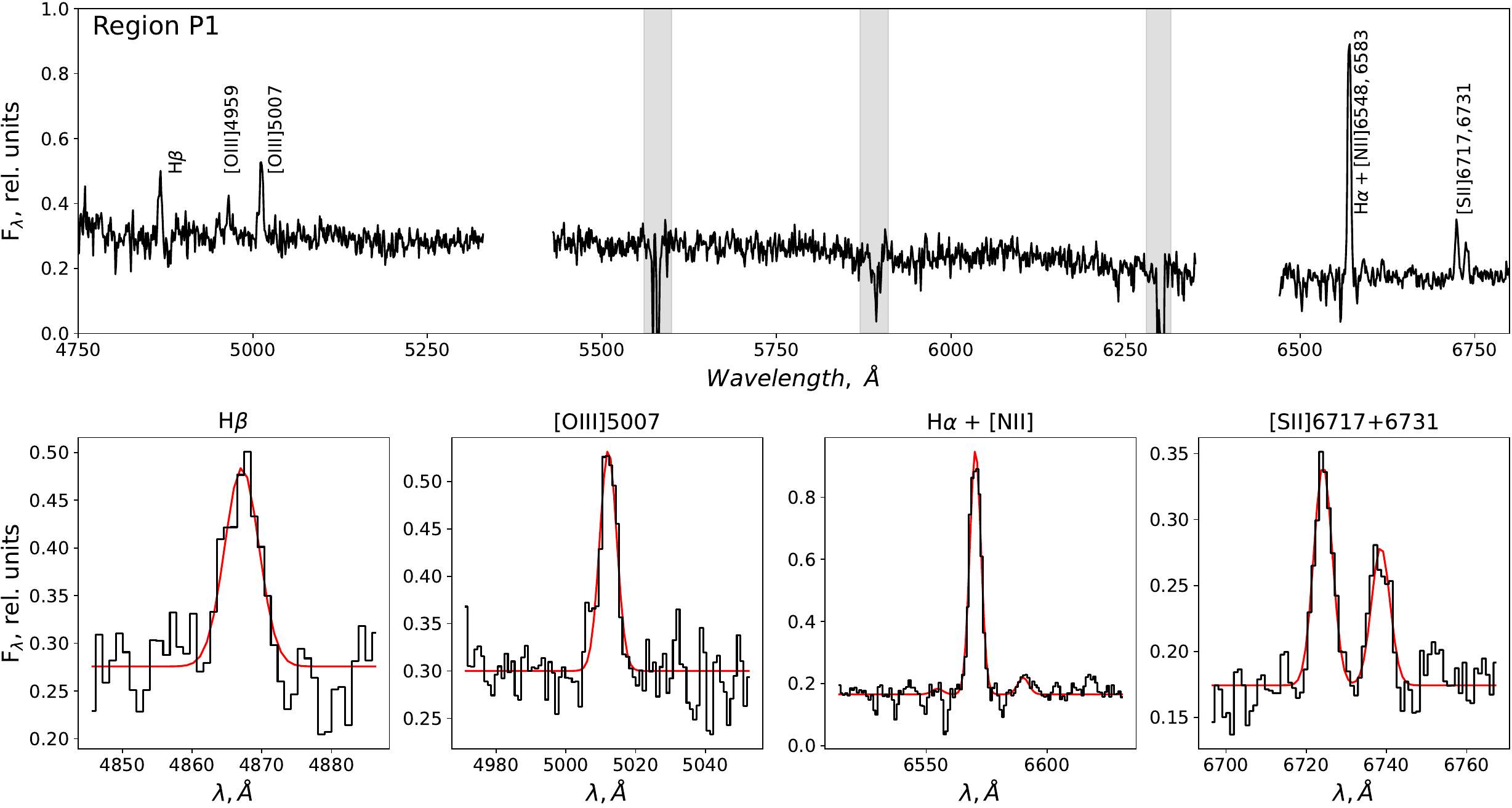}
    \caption{\revone{Same as in Fig.~\ref{fig:S1_spectrum} but for the P1 region.})}
    \label{fig:P1_spectrum}
\end{figure*}

\begin{figure*}
    \centering
    \includegraphics[width=\linewidth]{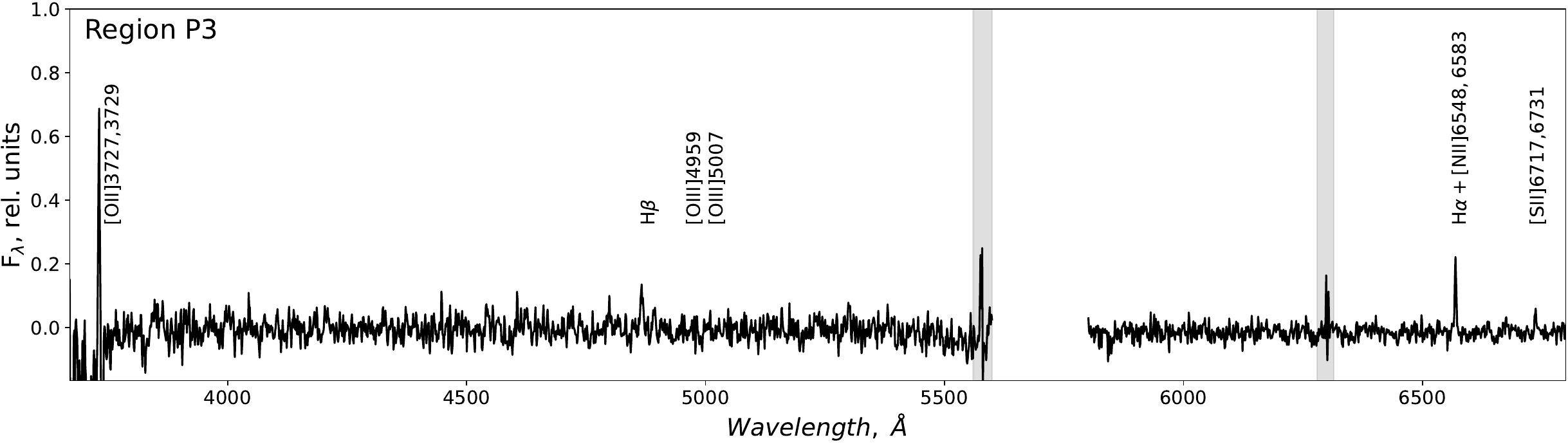}
    \caption{\revone{Optical spectrum for the P3 region obtained with TDS at CMO SAI MSU. Vertical grey stripes denote regions with bright sky line emission producing residual artefacts after their subtraction.}}
    \label{fig:P3_spectrum}
\end{figure*}

\revone{In this Section we present the spectra of individual regions with elevated \Ha velocity dispersion which we revealed in our analysis (Figs.~\ref{fig:S1_spectrum}--\ref{fig:P3_spectrum}). For each spectrum, we also provide the results of the best-fit approximation of the bright emission lines with a single Gaussian profile. The analysis of these spectra is presented in Section~\ref{sec:bpt}.}

\bsp	
\label{lastpage}
\end{document}